\documentclass[cernpreprint, atlasdraft=false, texlive=2023, UKenglish, texmf, orcidlogo]{atlasdoc}
 
\usepackage{atlaspackage}
\usepackage{atlasbiblatex}

\usepackage{atlasphysics}
 
\usepackage{tikz-feynman}
 
\usepackage{amsmath}

\graphicspath{{logos/}{figures/}}
 
\usepackage{ANA-TOPQ-2023-23-PAPER-defs}
 
\usepackage{array}
\newcolumntype{C}[1]{>{\centering}p{#1}}
\newcolumntype{R}[1]{>{\raggedleft}p{#1}}
\newcolumntype{L}[1]{>{\raggedright}p{#1}}

\AtlasTitle{Search for charged-lepton-flavour violating $\mu\tau qt$ interactions in top-quark production and decay in $pp$ collisions at $\sqrt{s}= 13$ TeV with the ATLAS detector at the LHC}

\AtlasAbstract{
A search for charged-lepton-flavour violating $\mu\tau qt$ ($q=u,c$) interactions is presented, considering both top-quark production and decay. The data analysed correspond to $140~\textrm{fb}^{-1}$ of proton--proton collisions at a centre-of-mass energy of $\sqrt{s}= 13$ TeV recorded with the ATLAS detector at the Large Hadron Collider.
The analysis targets events containing two muons with the same electric charge, a hadronically decaying $\tau$-lepton and at least one jet, with exactly one $b$-tagged jet, produced by a $\mu\tau qt$ interaction.
Agreement with the Standard Model expectation within $1.6$ standard deviations is observed, and limits are set at the $95\%$ confidence level (CL) on the charged-lepton-flavour violation branching ratio of $\mathcal{B}(t \to \mu\tau q) < 8.7 \times 10^{-7}$.
An Effective Field Theory interpretation is performed yielding $95\%$ CL limits on Wilson coefficients, dependent on the flavour of the associated light quark and the Lorentz structure of the coupling. These range from $|c_{\mathsf{lequ}}^{3(2313)}| / \Lambda^{2} < 0.10\textrm{ TeV}^{-2}$ for $\mu\tau ut$ to $|c_{\mathsf{ lequ}}^{1(2323)}| / \Lambda^{2} < 1.8\textrm{ TeV}^{-2}$ for $\mu\tau ct$. An additional interpretation is performed for scalar leptoquark production inducing charged lepton flavour violation, with fixed inter-generational couplings. Upper limits on leptoquark coupling strengths are set at the $95\%$ CL, ranging from $\lambda^{\textrm{LQ}} = 1.3$ to $\lambda^{\textrm{LQ}} = 3.7$ for leptoquark masses between $0.5$ and $2.0$ TeV.
}
 
\author{The ATLAS Collaboration}

\PreprintIdNumber{CERN-EP-2024-061}

\AtlasJournalRef{\PRD 110 (2024) 012014}
\AtlasDOI{10.1103/PhysRevD.110.012014}

\hypersetup{pdftitle={ATLAS document},pdfauthor={The ATLAS Collaboration}}
 
\begin{document}
 
\maketitle

% The next lines are included from the .//sections/Introduction.tex input file
\section{Introduction}
 
Prior to the prediction and subsequent observation of neutrino oscillations~\cite{Maki:1962mu,Pontecorvo:1967fh,PhysRevLett.81.1562,PhysRevLett.87.071301,PhysRevLett.89.011301}, the flavour of both charged and neutral leptons was assumed to be conserved within the Standard Model (SM). Extensions to the SM that can accommodate neutrino mass and mixing may also allow for the local non-conservation of charged-lepton flavour.
Charged-lepton-flavour violation (\LFV) may therefore occur at rates significantly lower than the current experimental sensitivity (for example, $\mu \rightarrow e\gamma$ due to a neutrino oscillation in a \Wboson-boson loop has a predicted branching ratio of $10^{-55}$~\cite{Calibbi:2017uvl}). Any observation of \LFV would hence provide strong evidence for new physics. Beyond-the-SM theories, such as those that predict leptoquarks (LQ), entail \LFV~\cite{Calibbi:2017uvl,LQreview} with a rate as high as $\mathcal{B}(t \to \ell\ell^{'} c) \approx 10^{-6}$~\cite{Kim:2018oih}, where $\ell,\ell^{'} = \{e, \mu, \tau\}$ and $\ell \neq \ell^{'}$. Examples include the unification of leptons and quarks into representations of the $\textrm{SU(5)}$~\cite{SU5}, $\textrm{SO(10)}$~\cite{SO10_1,SO10_2} or $\textrm{SU(5)}\otimes \textrm{U(1)}$~\cite{SU5U1_1,SU5U1_2,SU5U1_3,SU5U1_4} groups, supersymmetric scenarios~\cite{SUSY1,SUSY2,SUSY3,SUSY4,SUSY5} and technicolor models~\cite{TC1,TC2,TC3}. In addition, some deviations from SM predictions were observed in the comparisons of hadron decays involving $\tau$-leptons and other leptons that might hint at the presence of new phenomena, such as measurements of $R(D)$ and $R(D^{*})$~\cite{BaBar:2012obs,BaBar:2013mob,Belle:2015qfa,Belle:2016dyj,Belle:2017ilt,Belle:2019rba,LHCb:2023zxo,LHCb:2023uiv,Belle-II:2024ami,HFLAV:2022esi}.
 
Assuming that the energy scale probed experimentally is significantly lower than the scale of new physics, it is convenient to consider a model-independent approach with an effective field theory (EFT). Constraints on the operators describing two-quark two-lepton (2Q2L) contact interactions were compiled~\cite{Chala:2018agk}, and top-quark 2Q2L operators related to electrons and muons were probed by the CMS Collaboration at the Large Hadron Collider (LHC)~\cite{CMS:2023iul}. Top-quark 2Q2L operators involving muons and $\tau$-leptons however remain highly unconstrained~\cite{Chala:2018agk}, such that some \LFV top-quark interactions could be within the current sensitivity of the LHC~\cite{clfv}.
 
\begin{figure}[t!bph]
\centering
{\centering
\subfloat[]{\includegraphics[width=0.31\textwidth]{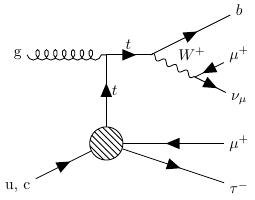}\label{fig:leptonic_production_tchannel_signal_feynman}}\hfill
\subfloat[]{\includegraphics[width=0.31\textwidth]{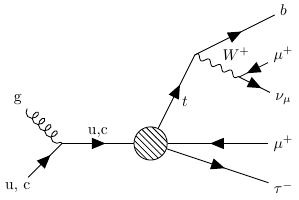}\label{fig:leptonic_production_schannel_signal_feynman}}\hfill
\subfloat[]{\includegraphics[width=0.33\textwidth]{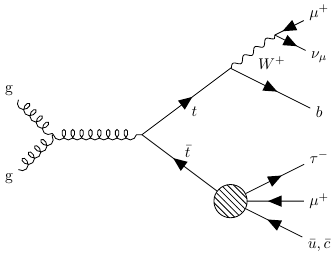}\label{fig:leptonic_decay_signal_feynman}}
}
\caption{Example Feynman diagrams of the process under study, where the hatched circle represents the cLFV vertex: \protect\subref{fig:leptonic_production_tchannel_signal_feynman} single-top-quark production with a cLFV process in the $t$-channel; \protect\subref{fig:leptonic_production_schannel_signal_feynman} single-top-quark production with a cLFV process in the $s$-channel; and \protect\subref{fig:leptonic_decay_signal_feynman} top-quark pair production with a cLFV top-quark decay process. The charge conjugate of each process is also considered.}
\label{fig:leptonic_signal_feynman}
\end{figure}

This analysis searches for the production of a single top quark via $gq_{k} \rightarrow t \ell{^\pm} \ell^{' \mp}$ and the charge-conjugate process, where $q_{k} = \{ u,c\}$ for $k = \{ 1,2 \}$ and $\ell\ell^{'} = \{\mu\tau, \tau\mu \}$.  The final state is chosen such that the top quark decays into a $b$-quark and a $W$ boson, and the $W$ boson subsequently decays leptonically into a muon and a neutrino. The $\tau$-lepton is required to decay hadronically. A cLFV top-quark decay in \ttbar events is also targeted, $t \to \ell^{\pm} \ell^{' \mp} q_{k}$ and the charge-conjugate process, where the other top quark of the pair decays into a muon according to the SM\@. 
These processes are shown in Figures~\ref{fig:leptonic_production_tchannel_signal_feynman} and~\ref{fig:leptonic_production_schannel_signal_feynman} for single-top quark production and Figure~\ref{fig:leptonic_decay_signal_feynman} for top-quark pair production with a cLFV top-quark decay process, each enabled through an EFT vertex.
The search presented is performed using the full Run 2 data sample of proton--proton ($pp$) collision data collected with the ATLAS detector in 2015--2018 at $\sqrt{s}= 13$ \TeV, corresponding to an integrated luminosity of \SI{140}{\per\fb}. The observed data are interpreted within the EFT framework, and also to test a leptoquark hypothesis~\cite{Schmaltz:2018nls}.
 
In the EFT framework, the $gq_{k} \rightarrow t \ell{^\pm} \ell^{' \mp}$ production and $t \to \ell^{\pm} \ell^{' \mp} q_{k}$ decay processes are described by the $\textrm{SU(3)}_{\textrm{C}}\otimes \textrm{SU(2)}_{\textrm{L}} \otimes \textrm{U(1)}_{\textrm{Y}}$ dimension-6 EFT operators listed in~\cref{tab:EFTop}. The list includes all the relevant 2Q2L operators that contribute, and consists of a subset of the Warsaw basis operators~\cite{Grzadkowski:2010es}. Wilson coefficients ($c$) may be assigned to each of the operators ($\mathcal{O}$) in~\cref{tab:EFTop}. The Wilson coefficients weight the contributions of the EFT operators in an effective Lagrangian, which assumes a mass scale of new physics, $\Lambda$, which is much larger than the energy scale directly accessible at the LHC:
 
\begin{equation}
\mathcal{L} = \mathcal{L}_{\mathrm{SM}} + \mathcal{L}_{\mathrm{eff}} = \mathcal{L}_{\mathrm{SM}} + \sum\limits_{\mathrm{x}} \frac{c_{\mathrm{x}}}{\Lambda^{2}}\mathcal{O_{\mathrm{x}}} + \cdots\,\, .
\end{equation}
 
\begin{table}[htpb]
\centering
\caption{EFT operator basis and degrees of freedom. In the convention used, $\mathsf{l}$ and $\mathsf{q}$ are the left-handed lepton and quark doublets, respectively, while $\mathsf{u}$ and $\mathsf{e}$ are the right-handed up-type quark and charged-lepton singlets, respectively. The indices $i,j = \{1,2,3\}$ represent the lepton flavour generations and $k,l = \{1,2,3\}$ represent the quark flavour generations, respectively. The Pauli matrices are denoted by $\sigma^{I}$, $\varepsilon=i\sigma^{2}$ is the antisymmetric $SU(2)$ tensor and $\sigma^{\mu\nu}=\frac{i}{2}[\gamma^{\mu},\gamma^{\nu}]$ and $\gamma^{\mu}$ are the Dirac matrices.}
\label{tab:EFTop}
\begin{tabular}{l|ll}
\toprule
Operator & Interaction & Lorentz Structure \\
\midrule
$\mathcal{O}_{\mathsf{lq}}^{1(ijkl)}$   & $(\bar{\mathsf{l}}_{i}\Gamu \mathsf{l}_{j})(\bar{\mathsf{q}}_{k}\gamu \mathsf{q}_{l})$ & Vector \\[5pt]
$\mathcal{O}_{\mathsf{lq}}^{3(ijkl)}$   & $(\bar{\mathsf{l}}_{i}\Gamu\sigma^{I}\mathsf{l}_{j})(\bar{\mathsf{q}}_{k}\gamu\sigma_{I}\mathsf{q}_{l})$ & Vector \\[5pt]
$\mathcal{O}_{\mathsf{eq}}^{(ijkl)}$    & $(\bar{\mathsf{e}}_{i}\Gamu \mathsf{e}_{j})(\bar{\mathsf{q}}_{k}\gamu \mathsf{q}_{l})$ & Vector \\[5pt]
$\mathcal{O}_{\mathsf{lu}}^{(ijkl)}$    & $(\bar{\mathsf{l}}_{i}\Gamu \mathsf{l}_{j}) (\bar{\mathsf{u}}_{k}\gamu \mathsf{u}_{l})$ & Vector  \\[5pt]
$\mathcal{O}_{\mathsf{eu}}^{(ijkl)}$    & $(\bar{\mathsf{e}}_{i}\Gamu \mathsf{e}_{j})(\bar{\mathsf{u}}_{k}\gamu \mathsf{u}_{l})$ & Vector  \\[5pt]
$\mathcal{O}_{\mathsf{lequ}}^{1(ijkl)}$ & $(\bar{\mathsf{l}}_{i}\mathsf{e}_{j})\varepsilon(\bar{\mathsf{q}}_{k} \mathsf{u}_{l})$ & Scalar \\[5pt]
$\mathcal{O}_{\mathsf{lequ}}^{3(ijkl)}$ & $(\bar{\mathsf{l}}_{i}\sigma^{\mu\nu}\mathsf{e}_{j})\varepsilon(\bar{\mathsf{q}}_{k}\sigma_{\mu\nu} \mathsf{u}_{l})$ & Tensor \\[5pt]
\bottomrule
\end{tabular}
\end{table}

The cLFV single-top-quark production cross-section grows quadratically with the values of the Wilson coefficients. Non-zero Wilson coefficients would lead to a large cLFV single-top-quark production cross-section and this process therefore dominates the search sensitivity. However, it is also instructive to relate the Wilson coefficients concisely through the top-quark decay width $\Gamma$ 
in terms of six degrees of freedom as follows~\cite{Chala:2018agk}:

\begin{align}\label{eq:master}
\Gamma(t\to\ell^{+}_{i}\ell^{-}_{j}q_{k}) = \frac{m_{t}}{6144\pi^{3}}\left(
\frac{m_{t}}{\Lambda}\right)^{4} \left\{
4|c_{\mathsf{lq}}^{-(ijk3)}|^{2} +
4|c_{\mathsf{eq}}^{(ijk3)}|^{2} +
4|c_{\mathsf{lu}}^{(ijk3)}|^{2} \nonumber \right. \\
\left. + ~4|c_{\mathsf{eu}}^{(ijk3)}|^{2} +
2|c_{\mathsf{lequ}}^{1(ijk3)}|^{2} +
96|c_{\mathsf{lequ}}^{3(ijk3)}|^{2} \right\}\,\, ,
\end{align}
 
where $c_{\mathsf{lq}}^{-(ijk3)}\equiv c_{\mathsf{lq}}^{1(ijk3)}-c_{\mathsf{lq}}^{3(ijk3)}$ is a combination used to contain the interactions of two up-type quarks with two charged leptons (the alternative sum $c_{\mathsf{lq}}^{+(ijk3)}\equiv c_{\mathsf{lq}}^{1(ijk3)}+c_{\mathsf{lq}}^{3(ijk3)}$ contains the interations of two up-type quarks and two neutrinos)~\cite{Durieux:2014xla} and $m_{t}$ is the mass of the top quark. Real values of the Wilson coefficients are assumed, which implies invariance with respect to the ordering of the leptons and quarks: $c^{(jilk)} = c^{(ijkl)}$.

The current constraints on the
Wilson coefficients come from a reinterpretation~\cite{Chala:2018agk} of a previous ATLAS flavour-changing neutral current (FCNC) search in the $tZq$ channel~\cite{TOPQ-2017-06}, based on a subset of $36~\textrm{fb}^{-1}$ of the data sample used in this analysis. These Wilson coefficient limits range from $|c_{\mathsf{lequ}}^{3(2313)}| / \Lambda^{2} < 3.4~\textrm{\TeV}^{-2}$ for $\mu\tau ut$ to $|c_{\mathsf{lequ}}^{1(2323)}| / \Lambda^{2} < 29~\textrm{\TeV}^{-2}$ for $\mu\tau ct$, and are dependent on the flavour of the associated light quark and the Lorentz structure of the coupling. The range of Wilson coefficient limits set by this reinterpretation corresponds to minimal and maximal branching-ratio limits of approximately $\mathcal{B}(t \to \mu\tau u) < 3.5 \times 10^{-5}$ and $\mathcal{B}(t \to \mu\tau c) < 3.0 \times 10^{-4}$ for the respective operators.
 
\begin{figure}[b!p]
\centering
{\centering
\subfloat[]{\includegraphics[width=0.33\textwidth]{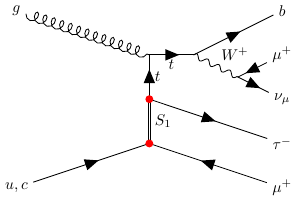}\label{fig:leptoquark_feynman_2}}\hfill
\subfloat[]{\includegraphics[width=0.33\textwidth]{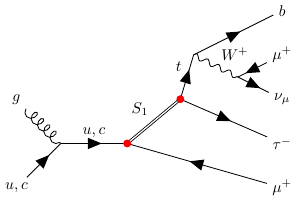}\label{fig:leptoquark_feynman_1}}\hfill
\subfloat[]{\includegraphics[width=0.33\textwidth]{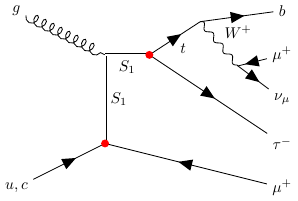}\label{fig:leptoquark_feynman_3}}
}
\caption{Example Feynman diagrams of the scalar leptoquark $S_{1}$ model under study, producing a single-top-quark via \protect\subref{fig:leptoquark_feynman_2} a non-resonant leptoquark, \protect\subref{fig:leptoquark_feynman_1} a resonant leptoquark and \protect\subref{fig:leptoquark_feynman_3} a resonant leptoquark with an off-shell leptoquark exchange. The leptoquark $S_{1}$ carries electric charge $Q = -\frac{1}{3}$, spin $S = 1$, lepton number $L = 1$ and baryon number $B = \frac{1}{3}$. 
In each case, initial states using up or charm quarks are assumed, and a final state consisting of two same-sign muons and one $\tau$-lepton is produced. The charge conjugate of each process is also considered. The dots represent vertices with a leptoquark-to-fermion coupling strength, $\lambda_{ki}$, for the appropriate quark generation ($k$) and lepton generation ($i$).}
\label{fig:leptoquark_signal_feynman}
\end{figure}

Leptoquarks, which arise in beyond-the-SM theories, are also a natural candidate for introducing cLFV interactions. The couplings between leptons and quarks may be inter-generational. In this interpretation, a scalar leptoquark ($S_{1}$) model is assumed that introduces couplings between all up-type quarks and all charged leptons, leading to the single leptoquark production processes presented in Figure~\ref{fig:leptoquark_signal_feynman} with the relevant final states.
 
This model, in which the leptoquark may couple to multiple generations of leptons and quarks, contains many degrees of freedom. The search for multi-generational single scalar leptoquarks is an extension to previous studies by ATLAS~\cite{EXOT-2022-39} and CMS~\cite{CMS:2023qdw} that search for single leptoquark production coupling to a single lepton and single quark generation only. 
The investigation of a fully general and complex scenario is beyond the scope of this measurement and for that reason, assumptions are made to simplify the multi-generational hypothesis.
The couplings of $S_{1}$ to each generation of quarks $k$ and leptons $i$, $\lambda_{ki}$, are fixed relative to one another such that the coupling of the leptoquark to the SM may be described by a single parameter $\lambda^{\textrm{LQ}}$:
 
\begin{equation}\label{eq:lqhierarchy}
\lambda_{ki} \in \:
\begin{pmatrix}
\lambda_{t\tau} & \lambda_{c\tau} & \lambda_{u\tau} \\
\lambda_{t\mu} & \lambda_{c\mu} & \lambda_{u\mu} \\
\lambda_{te} & \lambda_{ce} & \lambda_{ue}
\end{pmatrix} \equiv
\lambda^{\textrm{LQ}}
\begin{pmatrix}
10 & 1 & 0.1\\
1 & 0.1 & 0.01 \\
0.1 & 0.01 & 0.001
\end{pmatrix} \; ,
\end{equation}

where the largest value is the $S_{1}$-$t$-$\tau$ vertex and the smallest is the $S_{1}$-$u$-$e$ vertex.
A \textit{flavour hierarchy} in the coupling strengths is a common assumption in multi-generational leptoquark models. However, the magnitude of the hierarchies in the quark generations $(\lambda_{(k-1)i} / \lambda_{ki})$ and in the lepton generations $(\lambda_{k(i-1)} / \lambda_{ki})$ are not yet a matter of consensus, as different models span ratios ranging from $\sqrt{2}$ to $\frac{1}{16}$ for each~\cite{Alvarez:2018jfb,DiLuzio:2017vat,deMedeirosVarzielas:2019lgb,Chang:2016zll,Aebischer:2022oqe}.
In this search, a constant ratio in both the quark and lepton generations is assumed, $R = \lambda_{(k-1)i} / \lambda_{ki} = \lambda_{k(i-1)} / \lambda_{ki}$. An order of magnitude reduction ($R=0.1$) is chosen for each generational step down in either the quark or lepton flavour as a representative scenario, where $\lambda_{t\tau}$ is the strongest coupling. As $R$ affects the relative sizes of the $\lambda_{ki}$ couplings, this modifies the width of the leptoquark and the resulting kinematic properties of the decay. It is not straightforward to scale to alternative hierarchy assumptions and such a study is beyond the scope of this interpretation. However, this interpretation represents a robust first search for single scalar leptoquark production with large inter-generational couplings using the top quark.
The search for the production of a single scalar leptoquark $S_{1}$ is performed using the same analysis strategy as is optimised for the EFT interpretation. Limits on the cross-section of the $S_{1}$ model are expected to be slightly weaker than in the EFT interpretation due to this choice of optimisation.

% End of text imported from the .//sections/Introduction.tex input file

% The next lines are included from the .//sections/ATLASDetector.tex input file
\section{ATLAS detector}
 
\newcommand{\AtlasCoordFootnote}{
ATLAS uses a right-handed coordinate system with its origin at the nominal interaction point (IP) in the centre of the detector and the \(z\)-axis along the beam pipe. The \(x\)-axis points from the IP to the centre of the LHC ring, and the \(y\)-axis points upwards. Polar coordinates \((r,\phi)\) are used in the transverse plane, \(\phi\) being the azimuthal angle around the \(z\)-axis. The pseudorapidity is defined in terms of the polar angle \(\theta\) as \(\eta = -\ln \tan(\theta/2)\) and is equal to the rapidity $ y = \frac{1}{2} \ln \left( \frac{E + p_z c}{E - p_z c} \right) $ in the relativistic limit. Angular distance is measured in units of \(\Delta R \equiv \sqrt{(\Delta y)^{2} + (\Delta\phi)^{2}}\).}

The ATLAS detector~\cite{PERF-2007-01} at the LHC covers nearly the entire solid angle around the collision point.\footnote{\AtlasCoordFootnote}
It consists of an inner tracking detector surrounded by a thin superconducting solenoid, electromagnetic and hadronic calorimeters,
and a muon spectrometer incorporating three large superconducting air-core toroidal magnets.
 
The inner-detector system (ID) is immersed in a \qty{2}{\tesla} axial magnetic field
and provides charged-particle tracking in the range \(|\eta| < 2.5\).
The high-granularity silicon pixel detector covers the vertex region and typically provides four measurements per track,
the first hit generally being in the insertable B-layer (IBL) installed before Run~2~\cite{ATLAS-TDR-19,PIX-2018-001}.
It is followed by the SemiConductor Tracker (SCT), which usually provides eight measurements per track.
These silicon detectors are complemented by the transition radiation tracker (TRT),
which enables radially extended track reconstruction up to \(|\eta| = 2.0\).
The TRT also provides electron identification information
based on the fraction of hits (typically 30 in total) above a higher energy-deposit threshold corresponding to transition radiation.
 
The calorimeter system covers the pseudorapidity range \(|\eta| < 4.9\).
Within the region \(|\eta|< 3.2\), electromagnetic calorimetry is provided by barrel and
endcap high-granularity lead/liquid-argon (LAr) calorimeters,
with an additional thin LAr presampler covering \(|\eta| < 1.8\)
to correct for energy loss in material upstream of the calorimeters.
Hadronic calorimetry is provided by the steel/scintillator-tile calorimeter,
segmented into three barrel structures within \(|\eta| < 1.7\), and two copper/LAr hadronic endcap calorimeters.
The solid angle coverage is completed with forward copper/LAr and tungsten/LAr calorimeter modules
optimised for electromagnetic and hadronic energy measurements respectively.
 
The muon spectrometer (MS) comprises separate trigger and
high-precision tracking chambers measuring the deflection of muons in a magnetic field generated by the superconducting air-core toroidal magnets.
The field integral of the toroids ranges between \num{2.0} and \qty{6.0}{\tesla\metre}
across most of the detector.
Three layers of precision chambers, each consisting of layers of monitored drift tubes, cover the region \(|\eta| < 2.7\),
complemented by cathode-strip chambers in the forward region, where the background is highest.
The muon trigger system covers the range \(|\eta| < 2.4\) with resistive-plate chambers in the barrel, and thin-gap chambers in the endcap regions.
 
The luminosity is measured mainly by the LUCID--2~\cite{LUCID2} detector that records Cherenkov light produced in the quartz windows of photomultipliers located close to the beampipe.
 
Events are selected by the first-level trigger system implemented in custom hardware,
followed by selections made by algorithms implemented in software in the high-level trigger~\cite{TRIG-2016-01}.
The first-level trigger accepts events from the \qty{40}{\MHz} bunch crossings at a rate below \qty{100}{\kHz},
which the high-level trigger further reduces in order to record complete events to disk at about \qty{1}{\kHz}.
 
A software suite~\cite{ATL-SOFT-PUB-2021-001} is used in data simulation, in the reconstruction
and analysis of real and simulated data, in detector operations, and in the trigger and data acquisition
systems of the experiment.

% End of text imported from the .//sections/ATLASDetector.tex input file

% The next lines are included from the .//sections/AnalysisStrategy.tex input file
\section{Analysis strategy}
 
An overview of the analysis strategy is provided here, before each step is described in detail throughout the following sections.
 
A signal-enriched region (SR) is defined to target the signal cLFV processes and selects events that contain two muons, one hadronically decaying $\tau$-lepton and at least one jet. Exactly one jet is required to be identified as containing $b$-hadrons ($b$-tagged). The two muons are required to have the same electric charge (referred to as same sign (SS)), to avoid large background contributions of opposite-sign (OS) $\mu^{+}\mu^{-}$ pairs from $Z$-boson decays.
 
In addition, control regions (CRs) are defined that are enriched in backgrounds from fake or non-prompt (NP) muons or fake $\tau$-lepton candidates, and are depleted of signal events.
The main backgrounds stem from \ttbar events with NP muons from heavy-flavour hadron decays inside jets, associated top-quark production processes ($\ttbar Z, \ttbar W, \ttbar H$) with prompt leptons, diboson events ($WZ$, $ZZ$) with prompt leptons, and events with jets that are misidentified as hadronically decaying $\tau$-leptons (referred to as fake $\tau$-lepton events).
The prompt-lepton backgrounds are modelled by Monte Carlo (MC) simulations. The yield of the NP muon background is determined through a template fit in a dedicated CR, denoted $\textrm{CR}\ttbar\mu$.
This CR contains dilepton \ttbar events with one electron and one muon, and an additional NP muon passing looser identification and isolation requirements, mainly originating from the decay of a $B$-meson in a hadronic jet. The contribution of the fake $\tau$-lepton background is determined through a data-driven scale factor method in a dedicated CR, denoted $\textrm{CR}\tau$. This CR targets two OS muons and one hadronically decaying $\tau$-lepton and is designed to be enriched in events with a jet misidentified as a $\tau$-lepton.
The fake $\tau$-lepton scale factors are determined in $\textrm{CR}\tau$ as the first step of the analysis and propagated to the SR.
 
The signal contribution in the SR is then estimated with a binned profile-likelihood fit to the distribution of $H_{\mathrm{T}}$ (the scalar sum of the lepton and jet transverse momenta), with systematic uncertainties modelled as nuisance parameters. The region $\textrm{CR}\ttbar\mu$ is included in the fit, to simultaneously determine the normalisation of the NP muon background. For each EFT coupling contributing to the signal process, MC samples are generated separately for single-top-quark production and top-quark decay. Two inclusive samples are also generated with all couplings activated simultaneously. These inclusive samples are used together to determine an upper limit on the inclusive branching-ratio ($\mathcal{B}(t \rightarrow \mu\tau q)$) limit, and the separate EFT samples are used to determine limits on each operator individually (as specified in~\cref{tab:EFTop}). Finally, dedicated samples are generated for the leptoquark model and the fit is repeated to set limits on this hypothesis.

% End of text imported from the .//sections/AnalysisStrategy.tex input file

% The next lines are included from the .//sections/DataMC.tex input file
\section{Data and simulated event samples}
\label{sec:DataMC}
 
The $pp$ collision data analysed for this search were recorded with the ATLAS detector from 2015 to 2018 at a centre-of-mass energy of $\sqrt{s} = 13$~\TeV.
Events were selected using single-muon and single-electron triggers~\cite{TRIG-2018-05,TRIG-2018-01} and are required to have a
reconstructed primary vertex that has at least two associated tracks with transverse momentum (\pT) greater than 500~\MeV,
where the primary vertex is defined as that for which the associated tracks have the highest sum of $\pT^2$.
After the application of data-quality requirements~\cite{DAPR-2018-01}, the data sample corresponds to an integrated luminosity of 140~fb$^{-1}$, as determined using the LUCID-2 detector~\cite{LUCID2,DAPR-2021-01} for the primary luminosity measurements.
 
To optimise the event selection and to predict contributions from various SM processes, MC simulated event samples are used.
The effect of multiple interactions in the same and neighbouring bunch crossings (pile-up) was modelled by overlaying the simulated hard-scattering event with inelastic $pp$ events generated by \PYTHIA[8.186]~\cite{Sjostrand:2007gs} using the \NNPDF[2.3lo] set of parton distribution functions (PDFs)~\cite{Ball:2012cx} and parameter values set according to the A3 tune~\cite{ATL-PHYS-PUB-2016-017}.
After the event generation, the ATLAS detector response was simulated~\cite{SOFT-2010-01} with the \GEANT toolkit~\cite{Agostinelli:2002hh} with either the full simulation of the ATLAS detector or a fast-simulation package that relies on a parameterisation of the calorimeter response~\cite{ATL-PHYS-PUB-2010-013}.
In all processes, the top-quark mass was set to $172.5~\gev$.
All samples of simulated events, except those produced with the \SHERPA~\cite{Bothmann:2019yzt} generator, use \EVTGEN~\cite{Lange:2001uf}
to model the decays of bottom and charm hadrons. \SHERPA uses built-in decay tables from measurements.
 
Signal and background contributions to the search for cLFV are estimated by using MC event generators. Both cLFV single-top-quark production and top-quark pair production with a cLFV top-quark decay are treated as signal, while all other processes are treated as backgrounds.  These backgrounds include SM top-quark pair and single-top-quark production, the associated production of top quarks with \Wboson, \Zboson, $\gamma$ or Higgs bosons, and the production of diboson, triboson and \Wboson/\Zboson+jets. Possible contributions of the aforementioned EFT operators to the associated top quark production backgrounds are assumed to be negligible and thus disregarded.
 
The EFT signal processes are simulated as $gq_{k} \to t\ell\ell^{'}$ and $\ttbar \to (\ell\ell^{'}q_{k})((W\to \ell\nu) b)$
for the production and decay diagrams respectively, with $\ell, \ell^{'} = \{e, \mu, \tau\}$ and $q_{k} = \{u,c\}$. For this purpose a UFO model~\cite{ufo} containing the EFT operators listed in \cref{tab:EFTop} was created with \textsc{Feynrules 2.0}~\cite{feynrules} using the \textsc{dim6top} model~\cite{AguilarSaavedra:2018nen}. Events were generated at leading-order (LO) in QCD with \MADGRAPH{2.9.5}~\cite{Alwall:2014hca} for the hard process, in combination with \PYTHIA[8.306]~\cite{Bierlich:2022pfr} for showering and hadronisation. All the decay channels of the $\tau$-lepton are included. The five-flavour scheme is used, in which all the quark masses are set to zero except for the top quark. The renormalisation and factorisation scales (\muR, \muF) are dynamic and correspond to the centre-of-mass energy of the incoming partons in the case of the top-quark decay process and half the sum of the transverse masses of all final state particles and partons for the single-top-quark production process. The \NNPDF[3.1nlo]~\cite{NNPDF:2017mvq} PDF
was chosen; \PYTHIA{8} was configured according to the \textsc{A14} tune~\cite{ATL-PHYS-PUB-2014-021}. 
Theoretical cross-sections for single-top-quark production and top-quark decays through cLFV interactions are shown in Table~\ref{tab:clfv_xsecs_all}. The cross-sections for the \LFV \ttbar decay samples are evaluated through
 
\begin{equation}
\sigma_{\textrm{CLFV}} = 2\cdot\sigma_{\ttbar}\cdot\mathcal{B}(t\to W(\to\ell\nu)b)\cdot\frac{\Gamma(t\to\ell^{+}_{i}\ell^{-}_{j}q_{k})}{\Gamma_{t}},
\end{equation}
 
where $\sigma_{\ttbar}$ is the \ttbar production cross-section prediction at next-to-next-to-leading-order (NNLO) in QCD including the resummation of next-to-next-to-leading logarithmic (NNLL) soft-gluon terms calculated using \TOPpp[2.0]~\cite{Czakon:2011xx}. The relative scale and PDF uncertainties from the LO generation are maintained. The single-top-quark production cross-sections are determined using \MADGRAPH{2.9.5} with \textsc{dim6top} and these values and related kinematic distributions were checked against the \textsc{SMEFTsim3.0}~\cite{Brivio:2017btx,Brivio:2020onw} UFO model and were found to be compatible.
The cross-sections in Table~\ref{tab:clfv_xsecs_all} are the sum over all lepton flavour generations ($i,j = 1,2,3$ 
where $i \neq j$) and assume a top-quark mass of $172.5$ \GeV, an LO top-quark decay width $\Gamma_{t}$ of $1.51$ \GeV, a new physics scale of $\Lambda=1$ \TeV and a value of 1.0 for each Wilson coefficient.
The column labelled $c^{(ijk3)}_{\textrm{vector}}$ represents the individual cross-section contributions from each of $c_{\mathsf{lq}}^{-(ijk3)}$, $c_{\mathsf{eq}}^{(ijk3)}$, $c_{\mathsf{lu}}^{(ijk3)}$ and $c_{\mathsf{eu}}^{(ijk3)}$. While each vector coupling contributes comparably to the enhancement of the top-quark width, as shown in Eq.~(\ref{eq:master}), they may result in different kinematic distributions.
The production cross-sections are quoted for up and charm quark couplings separately ($k = 1,2$) and differ due to the proton PDFs\@. The \ttbar\ cross-sections involving $\ell\ell^{'}q_{k}$ decays are quoted for up and charm quarks together 
as these are not distinguished in the analysis.
 
To consider an uncertainty due to the parton shower generator, a second set of MC samples for the EFT signals are generated using the \HERWIG[7.1.6]~\cite{Bahr:2008pv,Bellm:2015jjp} prediction with the \HERWIG[7.1] default tune instead of the \PYTHIA[8.306] generator.

\begin{table}[htbp]
\centering
\caption{Theoretical cross-sections for single-top-quark production and top-quark decays through cLFV interactions involving vector, scalar and tensor EFT Wilson coefficients. The column titled as $c^{(ijk3)}_{\textrm{vector}}$ represents the individual cross-section contributions from each of $c_{\mathsf{lq}}^{-(ijk3)}$, $c_{\mathsf{eq}}^{(ijk3)}$, $c_{\mathsf{lu}}^{(ijk3)}$ and $c_{\mathsf{eu}}^{(ijk3)}$.
The coefficient indices represent the lepton flavour generations ($i,j = 1,2,3$ where $i \neq j$) and light quark flavour generations ($k = 1,2$). The single-top-quark production cross-sections are quoted for $u$- and $c$-quark couplings separately, while they are combined for the $\ttbar$ decay process ($q_k = u,c$). The scale and PDF uncertainties are given. The value of each Wilson coefficient is set to 1.0 for the calculation of the cross-section, with an energy scale of $\Lambda = 1$ TeV.}
\setlength\extrarowheight{5pt}
\small
\begin{tabular}{c|cccccccc}
\toprule
& \multicolumn{3}{c}{\textbf{Cross-section $\sigma^{+\textrm{scale}}_{-\textrm{scale}} \pm \textrm{PDF}$ [fb]}} \\
& $c^{(ijk3)}_{\textrm{vector}}$ & $c^{1(ijk3)}_{\mathsf{lequ}}$ & $c^{3(ijk3)}_{\mathsf{lequ}}$ \\ \hline
\textbf{Production $\ell\ell^{'} ut$} & $118^{+24}_{-19}\pm 1$ & $101^{+21}_{-16}\pm 1$ & $2150^{+410}_{-320}\pm 20$ \\
\textbf{Production $\ell\ell^{'} ct$} & $7.9^{+1.2}_{-1.0}\pm 1.6$ & $6.1^{+1.0}_{-0.8}\pm 1.5$ & $153^{+21}_{-18}\pm 29$ \\
\textbf{Decay $\ell\ell^{'} q_k t$} & $6.9^{+1.8}_{-1.3}\pm 0.1$ & $3.46^{+0.90}_{-0.66}\pm 0.03$ & $166^{+43}_{-32}\pm 2$ \\
\bottomrule
\end{tabular}
\label{tab:clfv_xsecs_all}
\end{table}
 
Alternative signal samples for a scalar leptoquark production, $gq_{k} \rightarrow S_{1} \ell$~($q_{k}=u,c$; $\ell=\mu,\tau$), and for $gq_{k} \to t\ell\ell^{'}$ with the $S_{1}$ exchanged as a virtual particle, were generated at LO in QCD with \MADGRAPH{2.9.5} for the hard process in combination with \PYTHIA[8.306] for showering and hadronisation and the \NNPDF[3.0nlo] PDF set~\cite{Ball:2014uwa}. The scales \muR and \muF correspond to the invariant mass of the leptoquark, $m_{S_{1}}$.
A dedicated UFO model~\cite{Diaz:2017lit,Schmaltz:2018nls} was utilised for this purpose, 
defining $S_{1}$ to carry electric charge $Q = -\frac{1}{3}$, spin $S = 1$, lepton number $L = 1$ and baryon number $B = \frac{1}{3}$. 
Samples were generated assuming a charm quark in the initial-state as the predominant production mechanism according to the flavour hierarchy described in Eq.~(\ref{eq:lqhierarchy}). 
Diagrams with an initial-state up quark contribute an additional $10\%$ to the cross-sections and were considered by reweighting the charm quark samples, accounting for small differences in the kinematics and acceptance. When following the leptoquark decay $S_{1} \rightarrow t \ell'$ $(\ell'=\tau,\mu)$, the total cross-section scales with $(\lambda^{\textrm{LQ}})^{4}$ as an effective contact interaction. The contribution to the expected sensitivity from initial-state up quarks is less than $3\%$.
The samples contain trilepton $\mu\mu\tau$ and $\tau\tau\mu$ final states only and cover a mass range of $0.5 < m_{S_{1}} < 2.5$ \TeV and a coupling range of $0.5 < \lambda^{\textrm{LQ}} < 3.5$. These samples are only used in Section~\ref{sec:LQSearch} of the paper, for the interpretation of the search in the leptoquark $S_{1}$ model. Processes involving a leptoquark in a top-quark decay ($t\rightarrow q_{k}\ell\ell^{'}$, mediated by $S_{1}$) are not considered as the contribution is expected to be negligible.
 
For the background processes, the production of \ttbar\ events was modelled using the \POWHEGBOX[v2]~\cite{Frixione:2007nw,Nason:2004rx,Frixione:2007vw,Alioli:2010xd} generator at next-to-leading-order (NLO) with the \NNPDF[3.0nlo] PDF set and the \hdamp\ parameter\footnote{The \hdamp\ parameter is a resummation damping factor and one of the parameters that controls the matching of \POWHEG matrix elements to the parton shower and thus effectively regulates the high-\pt\ radiation against which the \ttbar\ system recoils.} set to 1.5~times the top-quark mass~\cite{ATL-PHYS-PUB-2018-009}.
The events were interfaced to \PYTHIA[8.230]~\cite{Sjostrand:2014zea} to model the parton shower (PS), hadronisation, and underlying event, with parameters set according to the \textsc{A14} tune and using the \NNPDF[2.3lo] set of PDFs. The functional form of the renormalisation and factorisation scales was set to the default scale $m_\text{T}(t) = \sqrt{m_{t}^{2} + \pTX[2]}$, where $m_\text{T}(t)$ and $\pT$ are the transverse mass and transverse momentum of the top quark in each event.
The \ttbar\ sample is normalised to the cross-section prediction at NNLO in QCD including the resummation of NNLL soft-gluon terms calculated using \TOPpp[2.0].
 
The single-top-quark samples are split into the three processes: $s$-channel, $t$-channel and $tW$ associated production.
These samples were modelled using the \POWHEGBOX[v2] generator at NLO in QCD using the four-flavour (five-flavour) scheme for the $t$-channel ($s$-channel and $tW$)~\cite{Frederix:2012dh,Alioli:2009je} and the corresponding \NNPDF[3.0nlo] set of PDFs.
In the case of $tW$ production, the diagram-removal scheme~\cite{Frixione:2008yi} was used to address the interference with \ttbar production~\cite{ATL-PHYS-PUB-2018-009}. 
The events were interfaced with \PYTHIA[8.230], which used the \textsc{A14} set of tuned parameters and the \NNPDF[2.3lo] set of PDFs.
 
The associated production of a $\ttbar$ pair with a leptonically decaying \Zboson boson is modelled using the \MGNLO[2.8.1] generator, which provides matrix elements (ME) at NLO in QCD with the \NNPDF[3.0nlo]
PDF set. The functional form of \muR and \muF are set to the default scale $0.5 \times \sum_i \sqrt{m^2_i+p^2_{T,i}}$, where the sum runs over all the particles generated from the ME calculation.
Top-quark decays are simulated at LO using \MADSPIN~\cite{Frixione:2007zp,Artoisenet:2012st} to preserve all spin correlations.
The events are interfaced with \PYTHIA[8.244]
for the parton shower and hadronisation, using the \textsc{A14} set of tuned parameters
and the \NNPDF[2.3lo] PDF set.
 
The production of $\ttbar W$ events was simulated at NLO precision in QCD with \SHERPA[2.2.10] 
and the \NNPDF[3.0nnlo] PDF set~\cite{Ball:2014uwa}.
In this set-up, multiple MEs were matched and merged with the \SHERPA PS model based on the Catani--Seymour dipole factorisation scheme~\cite{Gleisberg:2008fv,Schumann:2007mg}.
The virtual QCD corrections (in the strong coupling constant $\alphas$) for MEs at NLO accuracy were provided by the \OPENLOOPS library~\cite{Buccioni:2019sur,Cascioli:2011va,Denner:2016kdg}. Up to one additional parton was included in
the NLO ME, and up to two additional partons were included at LO in QCD using Comix~\cite{Gleisberg:2008fv}. The merging scale parameter~(\muQ), which sets a threshold to determine what part of the phase-space
is filled by the PS or the ME generator, was set to an energy of \SI{30}{\GeV}.
In addition to the nominal prediction
at NLO in QCD, higher-order corrections related to electroweak (EWK) \ttW contributions (in the coupling $\alpha$) were also added as part of the signal definition. The $\alpha^{3}$ and $\alpha^{2} \alphas^{2}$ corrections
were added through MC event weights derived using the virtual additive corrections in the formalism described in Ref.~\cite{Kallweit:2015dum}. Second, sub-leading EWK corrections at order $\alpha^{3}\alphas$\,\cite{Frederix:2017wme} were
partially accounted for (only the real emission contribution) via the addition of an independent \Sherpa[2.2.10] sample produced at LO in QCD for this final state.
The combination of contributions of NLO QCD and NLO EWK effects taken from the \Sherpa configuration outlined above closely follows the strategy outlined in Ref.~\cite{Frederix:2021agh}.  This results in a total cross-section of $\sigma(\ttW) = 722\,\mathrm{fb}$, which was used for the normalisation of the simulation.
 
The production of \ttH events was modelled using the
\POWHEGBOX[v2]
generator, which provided matrix elements at NLO~\cite{Hartanto:2015uka} in QCD in the five-flavour scheme with the \NNPDF[3.0nlo]
PDF set. The functional forms of \muR and \muF were
set to \(\sqrt[3]{m_\text{T}(t)\cdot m_\text{T}(\bar{t}) \cdot m_\text{T}(H)}\).
The events were interfaced to \PYTHIA[8.230]
using the \textsc{A14} tune
and the \NNPDF[2.3lo]
PDF set.
The sample is normalised to the cross-section calculated at NLO QCD and NLO EWK accuracy~\cite{deFlorian:2016spz}.
 
The production of $tWZ$ events was modelled using the \MGNLO[2.3.3]
generator at NLO in QCD with the \NNPDF[3.0nlo] PDF set.
The events were interfaced with
\PYTHIA[8.212] 
using the \textsc{A14} tune
and the \NNPDF[2.3lo]
PDF set. The top quark and the \(Z\) boson decays were simulated at LO using \MADSPIN
to preserve spin correlations.
 
Samples of diboson final states (\(VV\)) were simulated with the
\SHERPA[2.2.1] ($WZ \rightarrow qqll, ZZ \rightarrow qqll$) or 2.2.2 (all other final states) generators, 
including off-shell effects and Higgs boson contributions, where appropriate.
Fully leptonic final states and semileptonic final states, where one boson
decays leptonically and the other hadronically, were simulated using
matrix elements at NLO accuracy in QCD for up to one additional parton
and at LO accuracy for up to three additional parton
emissions. Samples were generated using the \NNPDF[3.0nnlo] PDF set, 
along with the dedicated set of tuned parton-shower parameters developed by the
\SHERPA authors.
 
The production of \(V+\)jets was simulated with the \SHERPA[2.2.1]
generator. In this set-up, NLO matrix elements for up to two partons, and LO matrix elements for up
to four partons were calculated with the Comix 
and \OPENLOOPS libraries, and merged with the \SHERPA parton shower using the MEPS@NLO prescription~\cite{Hoeche:2012yf}, 
based on
Catani--Seymour dipole factorisation and the cluster hadronisation model~\cite{Winter:2003tt}. They employed the dedicated set of tuned parameters developed by the
\SHERPA authors and the \NNPDF[3.0nnlo] PDF set. 
 
The production of triboson (\(VVV\)) events was simulated with the
\SHERPA[2.2.2] 
generator using factorised gauge-boson decays.
Matrix elements, accurate to NLO for the inclusive process and to LO for up to
two additional parton emissions, were matched and merged with the \SHERPA parton
shower based on Catani--Seymour dipole factorisation 
using the \MEPSatNLO prescription~\cite{Hoeche:2011fd,Hoeche:2012yf,Catani:2001cc,Hoeche:2009rj}.
The virtual QCD corrections for matrix elements at NLO accuracy were
provided by the \OPENLOOPS library. 
Samples were generated using the \NNPDF[3.0nnlo] PDF set, 
along with the dedicated set of tuned parton-shower parameters developed by the \SHERPA authors.

Other processes make only minor contributions and are also modelled by MC: $t\bar{t}t\bar{t}$ is simulated with \MGNLO; $t\bar{t}t$ and $t\bar{t}WW$ with \MADGRAPH{2.2.2}; \ttgamma with \MADGRAPH{2.3.3}; $WH / ZH$ with \POWHEGBOX and $Z\gamma$ with \SHERPA 2.2.4. All samples are interfaced to \PYTHIA[8] except for the $Z\gamma$ sample.
An overlap removal procedure is applied to remove events in the \ttbar and \(Z+\)jets samples that have a photon in the matrix element, to prevent double-counting with the \ttgamma~and $Z\gamma$ samples.
The small contributions from \ttgamma~and $Z\gamma$, along with other processes that enter the event selection with a fake or non-prompt electron, are grouped and labelled `fake electron'.

% End of text imported from the .//sections/DataMC.tex input file

% The next lines are included from the .//sections/ObjectEventSelection.tex input file
\section{Object and event selection}
 
The events used in the analysis are selected with high efficiency using single-lepton triggers
, based on electron and muon signatures.
The lowest \pT thresholds were 20 \GeV for muons and 24 \GeV for electrons in 2015, and 26 \GeV for both lepton types in 2016--2018.  They were supplemented by additional triggers with higher \pT thresholds.
Events must contain at least one reconstructed lepton candidate corresponding to a lepton selected by the trigger (`trigger matched'), where the lepton \pT
exceeds the trigger \pT threshold by $1$ \GeV for electrons~\cite{TRIG-2018-05} or $5$\% for muons~\cite{TRIG-2018-01}.

Electron candidates are reconstructed from energy clusters (`clu') in the electromagnetic calorimeter that are associated with charged-particle tracks reconstructed in the inner detector~\cite{EGAM-2018-01}. Only candidates with $\pT > \SI{10}{\GeV}$ and $|\etaclu| < 2.47$ are considered. Candidates in the transition region between different electromagnetic calorimeter components, $1.37 < |\etaclu| <1.52$, are rejected. A multivariate likelihood discriminant combining shower-shape and track information is used to distinguish real electrons from hadronic showers. Isolation criteria, exploiting both calorimeter and tracking variables, and impact parameter requirements are used to reduce the background from non-prompt electrons produced in hadronic decays. 
Electron candidates are required to meet \emph{Tight} identification criteria based on the multivariate likelihood discriminant mentioned above~\cite{EGAM-2018-01}. 
The requirements for electrons on the transverse impact parameter significance and on the longitudinal impact parameter are $|d_{0}|/\sigma_{d_0} < 5$ and $| z_{0} \sin(\theta) | < \SI{0.5}{mm}$, respectively.
 
Muon candidates are reconstructed by combining inner detector tracks with track segments or full tracks in the muon spectrometer. 
Candidates are required to satisfy $\pt > \SI{10}{\GeV}$  and $|\eta| < 2.5$ and are required to satisfy a \emph{Medium}~\cite{MUON-2018-03} identification criterion.

Leptons from heavy-flavour hadron decays, misidentified jets, or photon conversions (collectively referred to as ‘non-prompt leptons’) are further suppressed using a boosted decision tree (BDT) discriminant, referred to as the non-prompt lepton BDT~\cite{MUON-2018-03}.
This BDT uses isolation and lifetime information associated with a track jet that matches the electron or muon candidate, and both tight and loose working points are defined.
Muon candidates populating the SR and $\textrm{CR}\tau$ must satisfy the non-prompt lepton BDT tight isolation requirement, with impact parameter selections of $|d_{0}|/\sigma_{d_0} < 3$ and $|z_{0} \sin(\theta) | < \SI{0.5}{mm}$; these are `Tight' muon candidates.
In $\textrm{CR}\ttbar\mu$, muons must satisfy the same longitudinal requirement of $|z_{0} \sin(\theta) | < \SI{0.5}{mm}$ but an inverted transverse impact parameter significance requirement of $|d_{0}|/\sigma_{d_0} \geq 3$ is applied, together with loose isolation requirements for the non-prompt lepton BDT\@.  These muon candidates are labelled `Loose'.

The constituents for jet reconstruction are identified by combining measurements from both the ID and the calorimeter using a \emph{Particle flow} algorithm~\cite{PERF-2015-09}. Jet candidates are reconstructed from these particle flow objects using the anti-$k_t$ algorithm~\cite{Cacciari:2008gp,Fastjet} with a radius parameter of $R=0.4$.  They are calibrated using simulation with corrections obtained from data using in situ techniques~\cite{JETM-2018-05}. Only jet candidates with  $\pT> 25$~\GeV\ and $|\eta|<2.5$ are selected.  To reduce the effect of pile-up,  each jet with $\pT<60$~\GeV\ and $|\eta|<2.4$ is required to satisfy the \emph{tight} working point of the jet vertex tagger (JVT)~\cite{PERF-2014-03}
criteria used to identify the jets as originating from the selected primary vertex. A set of quality criteria are also applied to reject events containing at least one jet arising from non-collision sources or detector noise~\cite{ATLAS-CONF-2015-029}.
 
Jets containing $b$-hadrons are identified ($b$-tagged) using the neural network `DL1r' algorithm~\cite{FTAG-2019-07}. The algorithm combines inputs from the impact parameters of displaced vertices and topological properties of secondary and tertiary vertices within a jet, and passes them to a neural network that outputs three values representing the probability of the jet being `light-flavour' (initiated by $u$-, $d$- or $s$-quarks or gluons), $c$- or $b$-jet, which is then combined into a single discriminant. The analysis uses a 77\% efficiency working point, calibrated using $\ttbar$ events at $\sqrt{s}=13$ \TeV, with rejection factors of 5.6 and 192 for $c$-jets and light-flavour jets, respectively~\cite{FTAG-2019-07}.
 
Hadronically decaying $\tau$-leptons, \tauhad, produce visible decay products along with a neutrino that escapes the detector undetected. The \tauhad candidates are seeded by jets reconstructed using the anti-$k_t$ algorithm
with a radius parameter $R = 0.4$ using topological energy clusters~\cite{PERF-2014-07} with a local hadronic calibration~\cite{ATL-LARG-PUB-2009-001}. Jets used to seed \tauhad reconstruction are required to have $\pt >\SI{10}{\GeV}$ and $|\eta| < 2.5$. All \tauhad candidates are required to have $\pt >\SI{20}{\GeV}$ and $|\eta| < 2.5$. As with electron candidates, \tauhad candidates in the transition region, $1.37 < |\eta| <1.52$, are rejected. The \tauhad production vertex is identified and a set of BDTs determines which of the reconstructed tracks are likely to originate from the hadronically decaying $\tau$-lepton~\cite{ATL-PHYS-PUB-2015-045}. The \tauhad candidates are required to have one or three associated tracks, referred to as 1-prong or 3-prong \tauhad, and these tracks must have measured charges summing to $\pm$1. A description of the $p_T$ calibration applied to \tauhad can be found in Ref.~\cite{ATLAS-CONF-2017-029}. To suppress \tauhad candidates originating from jets initiated by quarks or gluons, a recurrent neural network (RNN), described in Ref.~\cite{ATL-PHYS-PUB-2019-033}, is used to apply identification criteria. This makes use of information from individual tracks and calorimeter  clusters associated to the candidate along with high-level variables. An additional boosted decision tree discriminant, the eBDT, is used to reject \tauhad candidates originating from electrons. 
The \tauhad candidates are required to satisfy the `Medium' identification working point of the RNN~\cite{ATL-PHYS-PUB-2019-033}, corresponding to an efficiency of 75\% (60\%) for true 1-prong (3-prong) \tauhad.

To avoid double-counting of detector signatures, objects are removed in the following order:\footnote{For the overlap removal, $\Delta R$ is defined as $\Delta R \equiv \sqrt{\left(\Delta y\right)^2 + \left(\Delta \phi\right)^2}$, where $y$ is the rapidity of the object.}
electrons sharing a track with a muon;
jets within $\Delta R = 0.2$ of an electron;
electrons within $\Delta R = 0.4$ of a jet;
jets within $\Delta R = 0.4$ of a muon if they have at most two associated tracks;
muons within $\Delta R = 0.4$ of a jet;
\tauhad within $\Delta R = 0.2$ of an electron; 
\tauhad within $\Delta R = 0.2$ of a muon; 
any jet within $\dR = 0.2$ of the remaining \tauhad.
 
Scale factors (SFs) are used to correct the efficiencies in simulation to match the efficiencies measured in data for the electron~\cite{EGAM-2018-01,PERF-2017-03} and muon~\cite{MUON-2018-03,MUON-2022-01} trigger, reconstruction, identification and isolation criteria, as well as for the \tauhad reconstruction, identification and electron-rejection criteria~\cite{ATLAS-CONF-2017-029,ATL-PHYS-PUB-2019-033}.
Additional SFs are also applied for the JVT requirement~\cite{PERF-2014-03},
for pile-up and for the $b$-tagging efficiencies for jets that originate from the hadronisation of $b$-quarks~\cite{FTAG-2018-01}, $c$-quarks~\cite{FTAG-2020-08}, and $u$-, $d$-, $s$-quarks or gluons~\cite{FTAG-2019-02}.
 
Multiple selections are defined to create analysis regions that either focus on the cLFV signal processes, or constrain the normalisation of fake-lepton backgrounds.
All analysis regions are subject to the same lepton preselection requirements, shown in Table \ref{tab:mutau_regs}.
 
\begin{table}[htbp]
\centering
\caption{Common lepton selection requirements applied to all regions used in the analysis.}
\begin{tabular}{l|c}
\toprule
\textbf{Preselection:} & \\
Number of leptons & $N_{\ell} = 3, \ \pt > 10~\GeV, \ |\eta|<2.5$ \\
Leading muon / electron \pt & $\pt$ > 27~\GeV \\
Trigger matching & $\geq$ 1 trigger-matched muon / electron \\
Sum of lepton charges & $\sum q_{i} = \pm 1$ \\
\bottomrule
\end{tabular}
\label{tab:mutau_regs}
\end{table}
 
The SR requires two SS muons and a \tauhad candidate, at least one jet and exactly one $b$-tagged jet.
The SS muon requirement eliminates the large background from processes producing $\mu^{+}\mu^{-}$ pairs while the jet selection matches that of the signal process.
The $\textrm{CR}\tau$ region requires instead two OS muons and a \tauhad candidate, two or more jets and exactly one $b$-tagged jet. This region targets events with two prompt muons from \ttbar{} or \Zjets{}, in which a jet is misidentified as a \tauhad (denoted `fake $\tau$-lepton').
A requirement on the OS dimuon invariant mass in $\textrm{CR}\tau$ around the $Z$ boson mass is used to increase the proportion of light flavour jets counted as fake $\tau$-leptons (and decrease the contribution from heavy-flavour jets), consistent with the fraction observed in the SR.
 
The $\textrm{CR}\ttbar\mu$ region requires two muons and one electron of opposite charge to the highest \pt{} muon, at least one jet and at most two $b$-tagged jets.
Requiring that the two leptons with the highest \pT are an electron and a muon enhances the $t\bar{t}$ purity, as does the use of a veto on events with an OS dimuon invariant mass in the $Z$ boson mass window.
It also requires the lowest \pT lepton to be a Loose muon to preferentially select events with NP muons.
A selection on OS dilepton invariant masses and the leading muon \pt{} is used to reject signal events in $\textrm{CR}\ttbar\mu$ while retaining sufficient background statistics.
This requirement was derived in simulation and rejects signal events in which the $\tau$-lepton decays leptonically.
 
The full requirements for the analysis regions are reported in Table~\ref{tab:mutau_region_defs}.
NP muons tend to have lower \pt{} than those produced in the hard-scatter interaction.
The SR and $\textrm{CR}\tau$ regions therefore require both the muons to have \pt{} $>$ 15 \GeV to suppress this background while $\textrm{CR}\ttbar\mu$ retains low \pt{} muons.
 
For the single-top production (\ttbar decay) EFT process, 4.6\% (2.3\%) of generated MC events are reconstructed and satisfy the signal region event selection. Depending on the LQ mass and coupling strength, between 6.9\% and 8.1\% of generated MC events are reconstructed and satisfy the signal region event selection. These efficiencies are calculated relative to the number of generated events with $\mu\mu\tau$ final states.
 
\begin{table}[htbp]
\centering
\caption{Requirements for each analysis region. The symbol $\ell_{3}$ denotes the lowest \pt{} lepton. In $\mathrm{CR}\ttbar\mu$ an additional requirement is placed on the leading muon \pt{} ($p^{\mu_1}_\mathrm{T}$) and dilepton invariant masses in order to reject signal events.}
\begin{tabular}{c|c|c|c}
\toprule
& \textbf{SR} & \textbf{CR$\tau$} & \textbf{CR$\ttbar\mu$} \\
\midrule
Lepton flavour & \multicolumn{2}{c|}{2$\mu1$\tauhad} & $2\mu1e$ ~($\ell_{3} = \mu$) \\
$N_{\textrm{jets}}$ & $\geq$ 1 & $\geq$ 2 & $\geq$ 1 \\
$N_{b-\textrm{tags}}$ & 1 & 1 & $\leq 2$ \\
\tauhad \pT & $>20$ \GeV & $>20$ \GeV & -- \\
Muon \pT & $>15$ \GeV & $>15$ \GeV & $>10$ \GeV \\
Higher \pT muon & Tight & Tight & Tight \\
Lower  \pT muon & Tight & Tight & Loose \\
Muon charges & SS & OS & -- \\
$m^{\textrm{OS}}_{\mu\mu}$ & -- & -- & $>$15 \GeV \\
$|m^{\textrm{OS}}_{\mu\mu} - M_Z|$ & -- & $<$10 \GeV & $>$10 \GeV \\
$3p^{\mu_1}_\mathrm{T} + \sum m^{\textrm{OS}}_{\ell\ell}$ & -- & -- & $<$ 400 \GeV \\
\bottomrule
\end{tabular}
\label{tab:mutau_region_defs}
\end{table}

% End of text imported from the .//sections/ObjectEventSelection.tex input file

% The next lines are included from the .//sections/NPLepEstimation.tex input file
\newpage
 
\section{Non-prompt and fake-lepton estimations}
 
\subsection{Fake $\tau$-lepton estimation}\label{sec:nplepestimation_tau}
As \tauhad candidates are reconstructed from jets, some jets that did not originate from a $\tau$-lepton decay may be misidentified as a \tauhad. 
These fake $\tau$-leptons are suppressed by the \tauhad identification algorithm but still form a non-negligible background in the SR\@. To estimate the background from fake $\tau$-leptons, a SF method is used. The SF method involves applying multiplicative factors to the MC simulation to correct for mismodelling of the fake $\tau$-lepton background. 
The SFs are derived in $\textrm{CR}\tau$, for which the fake background in data is found by subtracting the non-fake background in MC from the data, and are determined separately for 1-prong and 3-prong \tauhad candidates.
These SFs are binned in the width\footnote{The jet width $w$ is defined as the \pt-weighted radial distance between objects associated to the jet and the jet axis: $w=\Sigma_j \Delta R^j \pt^j / \Sigma_j \pt^j$,  where $j$ denotes a jet constituent and $\Delta R^j$ is its radial distance from the centre of the jet.} of the jet from which the \tauhad is seeded,
and are derived from the data in $\textrm{CR}\tau$ as follows:
\begin{equation}
SF_{i} = \frac{N^{\textrm{CR}\tau}_{\textrm{Data},i} - N^{\textrm{CR}\tau}_{\textrm{MC},\textrm{non-fake},i}}{N^{\textrm{CR}\tau}_{\textrm{MC},\textrm{fake},i}} \, .
\end{equation}
The index $\textit{i}$ denotes a particular bin in the jet width distribution.
A parameterisation in the jet width is motivated as the SFs capture any mismodelling of the relative fractions of jets produced by quarks and gluons (quark-gluon fractions) for jets faking $\tau$-leptons, and any differences in the quark-gluon fractions between $\textrm{CR}\tau$ and the SR\@.
The estimated yield in the SR is then found by applying the SFs to the fake $\tau$-leptons in the MC:
\begin{equation}
N^{\textrm{SR}}_{\textrm{fakes},i} = N^{\textrm{SR}}_{\textrm{MC},\textrm{fake},i} \times SF_{i} \, .
\end{equation}
Sources of uncertainty relating to the prompt lepton background subtraction are accounted for by propagating all systematic uncertainties described in Section~\ref{sec:systematics} to the SF estimations, as well as uncertainties from limited MC statistics. 
Comparative SFs were also derived with alternative parameterisations, binned in jet \pT, jet $\eta$, missing transverse energy \met, total transverse energy $H_\mathrm{T}$ and in the separation of the \tauhad and muon, $\Delta R(\mu,\tauhad)$. In all cases, the fake $\tau$-lepton estimations in the SR were compatible.
 
\begin{figure}[b!p]
\centering
{\centering
\subfloat[]{\includegraphics[width=0.49\textwidth]{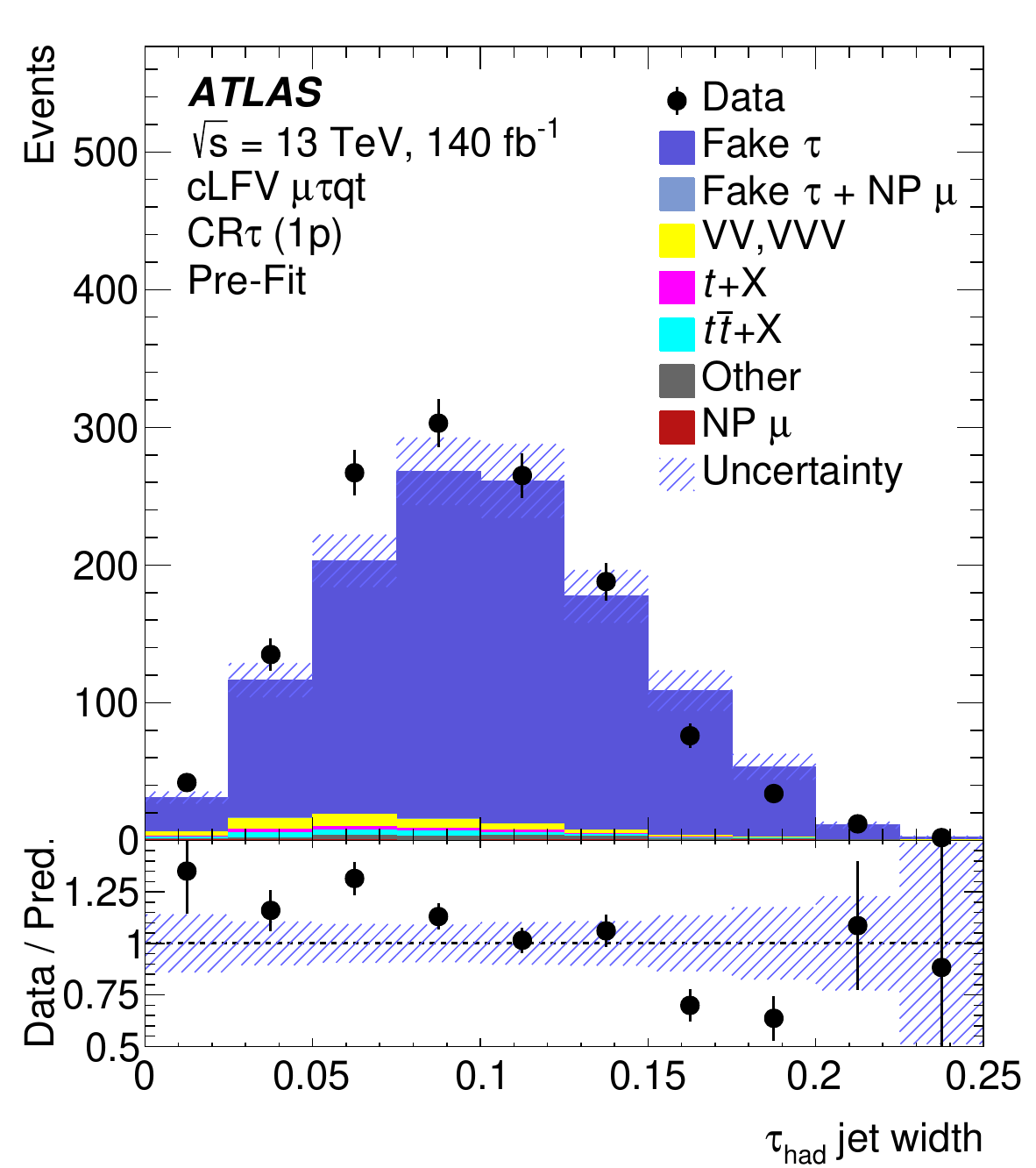}\label{fig:crtau_1p_plot}}
\subfloat[]{\includegraphics[width=0.49\textwidth]{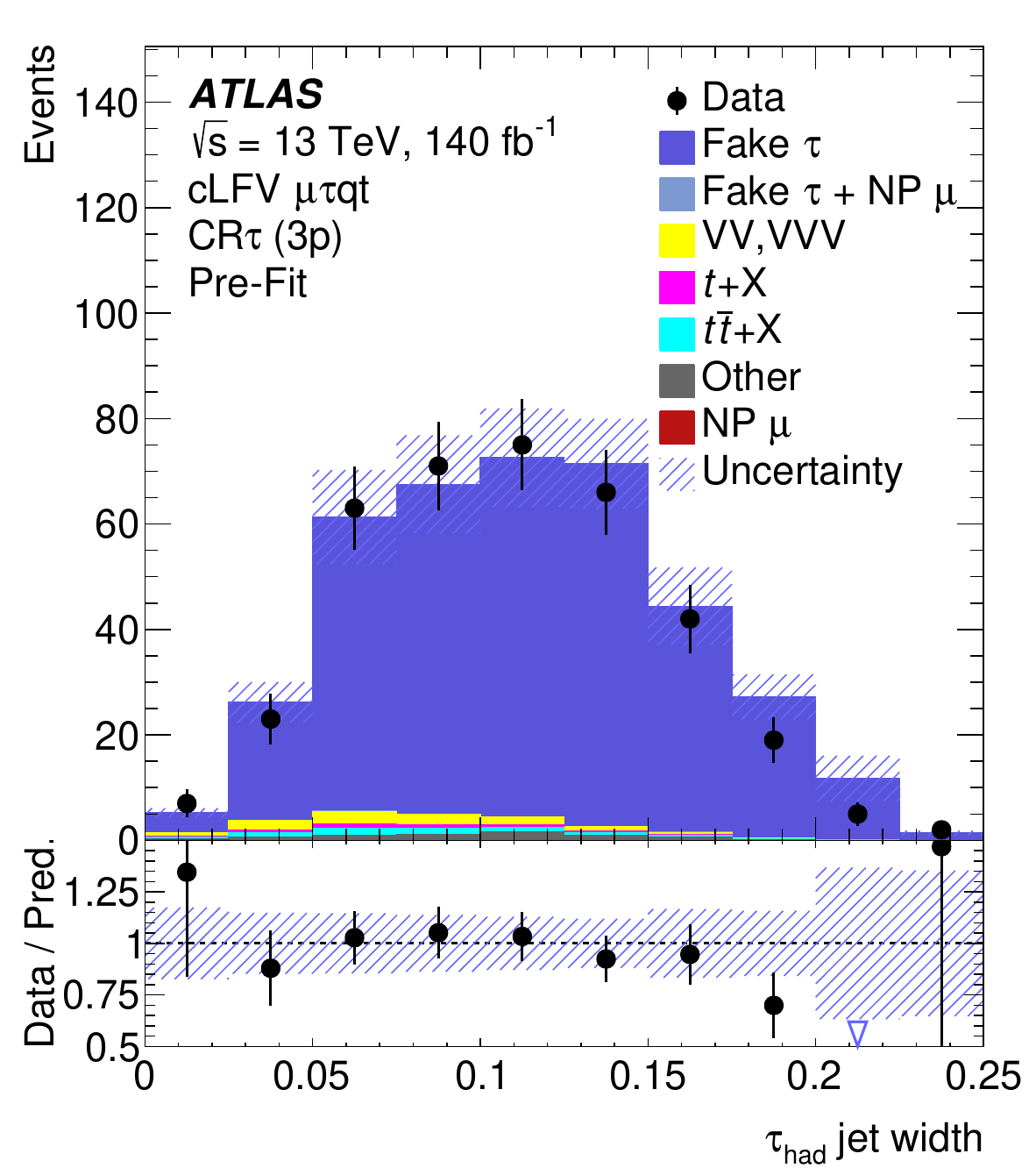}\label{fig:crtau_3p_plot}}
}
\caption{Observed distribution of the \tauhad jet width in $\textrm{CR}\tau$ compared with pre-fit expectations from Monte Carlo simulations, separated by \protect\subref{fig:crtau_1p_plot} 1-prong (1p) and \protect\subref{fig:crtau_3p_plot} 3-prong (3p) \tauhad track multiplicity. This region is used to derive scale factors for the data-driven fake $\tau$-lepton estimation. The legend entry of $\ttbar+X$ is the sum of $\ttbar W, \ttbar Z$ and $\ttbar H$. The lower panels show the ratio of the data (`Data') to the background prediction (`Pred.').  The uncertainty bands include the statistical and systematic uncertainties in the background prediction. An arrow on the lower panel indicates a point outside of the axis range. The last bin includes overflow events.}
\label{fig:mutau_control_plots_CRtau}
\end{figure}

Observed event yields in $\textrm{CR}\tau$ compared with pre-fit expectations from MC simulations are shown in Figure~\ref{fig:mutau_control_plots_CRtau} for 1-prong and 3-prong \tauhad candidates, as a function of the jet width,
prior to the calculation of the fake $\tau$-lepton SFs. The data and MC event yields in this region are given in Table~\ref{tab:mutau_region_yields_CRtau}.
 
\begin{table}[htbp]
\centering
\caption{Observed and predicted event yields for CR$\tau$ from Monte Carlo simulations with all systematic uncertainties, prior to the calculation of the fake $\tau$-lepton scale factors, separated by 1-prong (1p) or 3-prong (3p) \tauhad track multiplicity.}
\small
\begin{tabular}{c|S[table-format=4.4]cS[table-format=2.4]|S[table-format=3.3]cS[table-format=2.3]}
\toprule
\textbf{Process} & \multicolumn{3}{|c|}{\textbf{$\textrm{CR}\tau$ (1p)}} & \multicolumn{3}{|c}{\textbf{$\textrm{CR}\tau$ (3p)}}  \\
\midrule
Fake $\tau$ & 1150 & $\pm$ & 80 & 364 & $\pm$ & 28 \\
Fake $\tau$~+~NP~$\mu$ & 1.6 & $\pm$ & 1.2 & 0.3 & $\pm$ & 0.5 \\
WZ     & 22 & $\pm$ & 7 & 6.5 & $\pm$ & 2.0 \\
ZZ     & 11 & $\pm$ & 4 & 3.1 & $\pm$ & 1.0 \\
$t$+X    & 12.0 & $\pm$ & 0.9 & 3.41 & $\pm$ & 0.28 \\
\ttZ   & 16 & $\pm$ & 7 & 4.8 & $\pm$ & 2.3 \\
\ttW   & 0.65 & $\pm$ & 0.34 & 0.25 & $\pm$ & 0.15 \\
\ttH   & 0.84 & $\pm$ & 0.12 & 0.26 & $\pm$ & 0.04 \\
VVV    & 0.12 & $\pm$ & 0.06 & 0.027 & $\pm$ & 0.014 \\
\ttbar+~NP$~\mu$ & 1.0 & $\pm$ & 0.8 & 0.16 & $\pm$ & 0.24 \\
$Z$~+~NP~$\mu$  & 0.022 & $\pm$ & 0.009 & & -- & \\
Other  & 17 & $\pm$ & 9 & 6.1 & $\pm$ & 3.1 \\
\midrule
Total  & 1230 & $\pm$ & 90 & 389 & $\pm$ & 30 \\
\midrule
Data   & 1324 & & & 373 & &  \\
\bottomrule
\end{tabular}
\label{tab:mutau_region_yields_CRtau}
\end{table}
 
\subsection{Non-prompt muon estimation}
 
The dominant background in the SR is due to \ttbar events reconstructed with three leptons due to a NP muon from a heavy-flavour hadron decay inside a jet, typically a $b$-jet. A tight muon isolation working point is chosen to suppress this background but this process still constitutes a significant fraction of events that satisfy the event selection. A data-driven template fit is performed using $\textrm{CR}\ttbar\mu$ to fit the normalisation of the NP muon background. This method allows the data in the CR to correct the rate at which events with NP muons enter the fit. This background is therefore determined simultaneously with the limits set on the signal strength of the process under consideration. 
 
Events with both a NP muon and a fake $\tau$-lepton are corrected using the SFs described in \cref{sec:nplepestimation_tau} before being included in the NP muon estimation.

% End of text imported from the .//sections/NPLepEstimation.tex input file

% The next lines are included from the .//sections/Systematics.tex input file
\section{Systematic uncertainties}
\label{sec:systematics}
Systematic effects may change the expected numbers of events from the signal and background processes and the shape of the fitted distributions in the SR and the CRs.
These effects are evaluated by varying each source of systematic uncertainty by $\pm1\sigma$ and considering the resulting deviation from the nominal expectation as the uncertainty.
 
Uncertainties due to the modelling of both the EFT and leptoquark signals are estimated by considering independent variations of the renormalisation and factorisation scales by factors of 2 and 0.5, but normalising the signal to the nominal cross-section.
The uncertainties due to the PDF are calculated using the \NNPDF[3.1nlo] MC replica set. The standard deviation of this set is used to define a symmetrised uncertainty on the signal sample.
The uncertainty arising from the modelling of initial-state radiation (ISR) is evaluated by varying $\alpha_s$ in the initial-state shower, while that arising from final-state radiation (FSR) modelling is evaluated using variations of $\mu_{R}$ for QCD emissions in the final state shower.
 
Additionally, an uncertainty due to the choice of parton shower generator is estimated for the EFT signal by considering the change when using an alternative MC sample that uses the \HERWIG[7.1.6] 
prediction 
instead of the \PYTHIA[8.306] generator.
For $gq_{k} \rightarrow t \ell{^\pm} \ell^{' \mp}$ signal samples, \HERWIG is found to predict inaccurate branching fractions for $\tau$-lepton decays.
The branching fractions in these \HERWIG samples are subsequently corrected to match those predicted by \PYTHIA, before the evaluation of the parton shower uncertainty. The parton shower uncertainty is increased by applying a conservative uncertainty to the branching ratio corrections, based on the statistical uncertainty of the samples from which the corrections are derived.
 
Cross-section uncertainties in SM predictions are applied to each background process. A cross-section uncertainty of $6.1\%$ is used for the MC prediction of $\ttbar$ production~\cite{Czakon:2011xx}. 
The primary contribution of this process after the event selections is through NP muons, which may not be well modelled by the MC, and the overall normalisation is therefore determined in the fit.
For the single top process, a cross-section uncertainty of $5.3\%$ is included~\cite{Kidonakis:2010ux,Kidonakis:2013zqa}.
Cross-section uncertainties of $12\%$ and $10\%$, respectively, are applied to the \ttZ~and \ttH~processes, based on calculations reported in Ref.~\cite{deFlorian:2016spz}.
Some processes show discrepancies between data and MC in previous measurements; in these cases more conservative uncertainties are applied to cover the difference.  These include an uncertainty of $50\%$ for $\ttbar W$ production~\cite{TOPQ-2021-08,CMS-TOP-21-011,ATLAS:2024moy}
and $30\%$ for diboson production~\cite{TOPQ-2019-06,STDM-2018-03,HIGG-2016-29}.
For processes with very minor contributions to the signal regions, conservative uncertainties are applied of $40\%$ for $W$\!+jets and $Z$+jets production~\cite{TOPQ-2019-06}, $50\%$ for triboson production~\cite{TOPQ-2016-11} and $50\%$ for all other processes. The measurement is not sensitive to these minor contributions. 
 
For the background processes, uncertainties due to the renormalisation and factorisation scales are estimated separately for each process, following the same procedure as for the signal.
 
The uncertainties due to the PDF for the \ttbar, single-top-quark, \ttW~and \ttH~processes are calculated using the \PDFforLHC[15] prescription~\cite{Butterworth:2015oua} using the Hessian sets consisting of 30 eigenvectors. Each of the 30 eigenvectors is assigned a nuisance parameter that is profiled in the fit. For the \ttZ, \Zjets, \Wjets~and diboson background processes the uncertainties due to the PDF are estimated following the same procedure as for the signal, separately for each sample.
For the \ttbar~process, an additional uncertainty arising from the value of $\alpha_{s}(M_{Z}) = 0.118$, used to evaluate the PDF, is estimated by using the nominal PDF set evaluated with $\alpha_{s}(M_{Z}) = 0.117$ and $\alpha_{s}(M_{Z}) = 0.119$. This uncertainty has a negligible impact and is thus neglected for the subdominant background processes.
 
For the \ttbar process, the uncertainty in the matrix-element matching is assessed as the full observed difference between simulated samples generated with the \POWHEGBOX and \MGNLO programs. Both of the samples are showered with \Herwig[7.1.3]~\cite{Bahr:2008pv,Bellm:2015jjp},
using the \Herwig[7.1] default set of tuned parameters~\cite{Bellm:2015jjp,Bellm:2017jjp}
and the \MMHT[lo] PDF set~\cite{Harland-Lang:2014zoa}.
 
To cover the choice of parton shower and hadronisation model, samples are produced with the \POWHEGBOX program
and then showered with either \PYTHIA[8.230]
using the A14 tune 
and the \NNPDF[2.3lo] PDF set, or \Herwig[7.1.3]
using the \Herwig[7.1] default set of tuned parameters 
and the \MMHT[lo] PDF set. 
The difference is taken as a systematic uncertainty. The uncertainties due to the choice of parton shower generator for the $\ttbar H$ and $\ttbar Z$ backgrounds are also estimated by comparing the nominal predictions with an alternative set using \HERWIG[7.0.4]~\cite{Bahr:2008pv,Bellm:2015jjp} showering, with the H7UE set of tuned parameters~\cite{Bellm:2015jjp} and the \MMHT[lo] PDF set.
 
The $\ttbar W$ background has an uncertainty to cover the choice of generator. This is evaluated for the QCD component as the difference between a sample generated with \SHERPA[2.2.10] and an alternative sample using \MGNLO{}+\PYTHIA[8.245] with an FxFx merging scheme~\cite{Frederix:2012ps,Catani:2001cc,Hamilton:2012np}. To evaluate the generator uncertainty on the `subleading' EWK component, the \SHERPA[2.2.10] EWK sample is compared with a LO \MADGRAPH{2.6.2}+\PYTHIA{8} sample that also models these EWK effects.
 
For the \ttbar, single-top-quark, $\ttbar Z$ and $\ttbar H$ processes, the uncertainty arising from the modelling of ISR is evaluated using an independent variation of the A14 tune to its \texttt{Var3cUp} and \texttt{Var3cDown} variants~\cite{ATL-PHYS-PUB-2014-021}, corresponding to variations of $\alpha_s$ in the initial-state shower. For the \ttbar, single-top-quark and $\ttbar H$ processes the modelling of FSR is evaluated from an independent variation of the renormalisation scale for emissions from the parton shower by factors of 2.0 and 0.5.
For the \ttbar process, the uncertainty arising from the choice of the $h_{\textrm{damp}}$ parameter within \POWHEG is evaluated using an alternative \POWHEG+\Pythia[8] \ttbar sample in which $h_{\textrm{damp}} = 3.0\, m_{\mathrm{top}} $ (the default setting is $h_{\textrm{damp}} = 1.5\, m_{\mathrm{top}} $).

For the $tW$ single-top-quark process, an uncertainty arising from the removal of the overlap with the \ttbar\ process is estimated by replacing the diagram-removal scheme with the diagram-subtraction scheme~\cite{Frixione:2008yi}.

The uncertainty in the combined 2015--2018 integrated luminosity is 0.83\%~\cite{DAPR-2021-01}, obtained using the LUCID-2 detector~\cite{LUCID2} for the primary luminosity measurements, complemented by measurements employing the inner detector and calorimeters.
The MC samples were reweighted to reproduce the pile-up distributions in the data separately for the three periods 2015--2016, 2017 and 2018. The uncertainty due to pile-up is evaluated by varying the correction factors used to perform the reweighting within their uncertainties.

The jet energy scale uncertainty is derived by combining information from test-beam data,
LHC collision data and simulation, and the jet energy resolution uncertainty
is obtained by combining dijet \pt-balance measurements and simulation~\cite{JETM-2018-05}.
Additional considerations related to jet flavour, pile-up corrections,
$\eta$ intercalibration and high-\pt jets are included.
A total of 31 independent contributions are considered for the jet energy scale uncertainty, and 13 for the jet energy resolution uncertainty.
There is an additional uncertainty in the efficiency of the JVT algorithm to identify and remove jets from pile-up, which is measured with $Z \to \mu^+\mu^-$ events in data using techniques similar to those used in Ref.~\cite{PERF-2014-03}.
 
The efficiency to correctly tag $b$-jets is measured using dileptonic \ttbar\ events~\cite{FTAG-2018-01}. 
The mis-tag rate for $c$-jets is measured using semileptonic \ttbar\ events,
exploiting the $c$-jets from the hadronic \Wboson-boson decays
using techniques similar to those in Ref.~\cite{FTAG-2020-08}. 
The mis-tag rate for light-jets is measured using a negative-tag method,
similar to that in Ref.~\cite{FTAG-2019-02}, applied to \Zjets events.
Uncertainties on this efficiency and these mis-tag rates are due to reconstructed object calibrations
and to the modelling of the different processes, and are decomposed into sets of uncorrelated sources
of uncertainty: 9, 4, and 4 components for $b$-, $c$- and light-jets, respectively.
 
Uncertainties associated with electrons and muons arise from the trigger, reconstruction,
identification and isolation efficiencies, as well as the momentum scale and resolution.
These are measured using $Z\to \ell^+\ell^-$ and $J/\psi \to \ell^+\ell^-$ events ($\ell = e,\mu$)~\cite{EGAM-2018-01,MUON-2018-03}.
Uncertainties associated with \tauhad candidates cover reconstruction, identification and electron-rejection efficiencies, as well as the momentum correction, and were estimated by using $Z \rightarrow \tau^{+}\tau^{-}$, $Z \rightarrow e^{+}e^{-}$ and dijet events~\cite{ATLAS-CONF-2017-029,ATL-PHYS-PUB-2019-033,EXOT-2014-05}.

The uncertainty originating from the limited number of simulated MC events is implemented via the Barlow--Beeston approach~\cite{Barlow:249779}.
Three uncertainties for each bin of the fitted distributions are considered: one for the uncertainty originating from the SM backgrounds and one for each of the cLFV signals (single-top-quark production and top-quark pair production with a cLFV top-quark decay process).
 
The NP muon estimation is performed simultaneously with the limit-setting procedure so sources of uncertainty are accounted for in the same manner as the prompt backgrounds.
For the fake $\tau$-lepton estimation, each source of uncertainty is propagated through the scale factor method to provide systematic variations for the fake $\tau$-lepton component of the SR\@. The uncertainty arising from the limited data and MC statistics used to derive the SFs in $\textrm{CR}\tau$ is also considered.

% End of text imported from the .//sections/Systematics.tex input file

\newpage

% The next lines are included from the .//sections/Results.tex input file
\section{Results}
 
\subsection{Effective Field Theory Interpretation}
\label{sec:eft_search}
 
The normalisation factors of the EFT signal and the NP muon contributions are obtained from a simultaneous profile-likelihood fit to the SR and $\textrm{CR}\ttbar\mu$ with systematic uncertainties included as nuisance parameters. In both of the regions, the distribution of the scalar sum of the lepton and jet transverse momenta, $H_{\mathrm{T}}$, is used. This distribution  separates the EFT signal from the SM backgrounds, as the cLFV single-top-quark production diagrams (Figure~\ref{fig:leptonic_signal_feynman}) produce high-\pT leptons.
The likelihood function is constructed as a product of Poisson probability terms over all bins considered in the search. This function depends on a signal strength parameter, $\mu_\textrm{cLFV}$, which is defined as a multiplicative factor applied to the predicted yield for the two signal cLFV processes, the normalisation of the non-prompt muon contribution $k(\text{NP}\mu)$, and a set of nuisance parameters, $\theta$, that encode the effect of systematic uncertainties on the signal and background expectations. The expected total number of events in a given bin depends on $\mu_\textrm{cLFV}$, $k(\text{NP}\mu)$ and $\theta$.  All nuisance parameters are subject to Gaussian distribution constraints in the likelihood. For a given value of $\mu_\textrm{cLFV}$, the nuisance parameters $\theta$ allow variations of the expectations for signal and background according to the corresponding systematic uncertainties, and their fitted values result in the deviations from the nominal expectations that globally provide the best fit to the data. This procedure allows a reduction of the impact of systematic uncertainties on the search sensitivity. Statistical uncertainties in each bin are taken into account by dedicated parameters in the fit.  The best fit is obtained by performing a profile-likelihood fit to the data under the signal-plus-background hypothesis, maximising the likelihood function over $\mu_\textrm{cLFV}$, $k(\text{NP}\mu)$ and $\theta$. The fit is evaluated with the {\textsc RooFit} package~\cite{RooFit,RooFitManual}.
 
\begin{figure}[b!t!p]
\centering
{\centering
\subfloat[]{\includegraphics[width=0.50\textwidth]{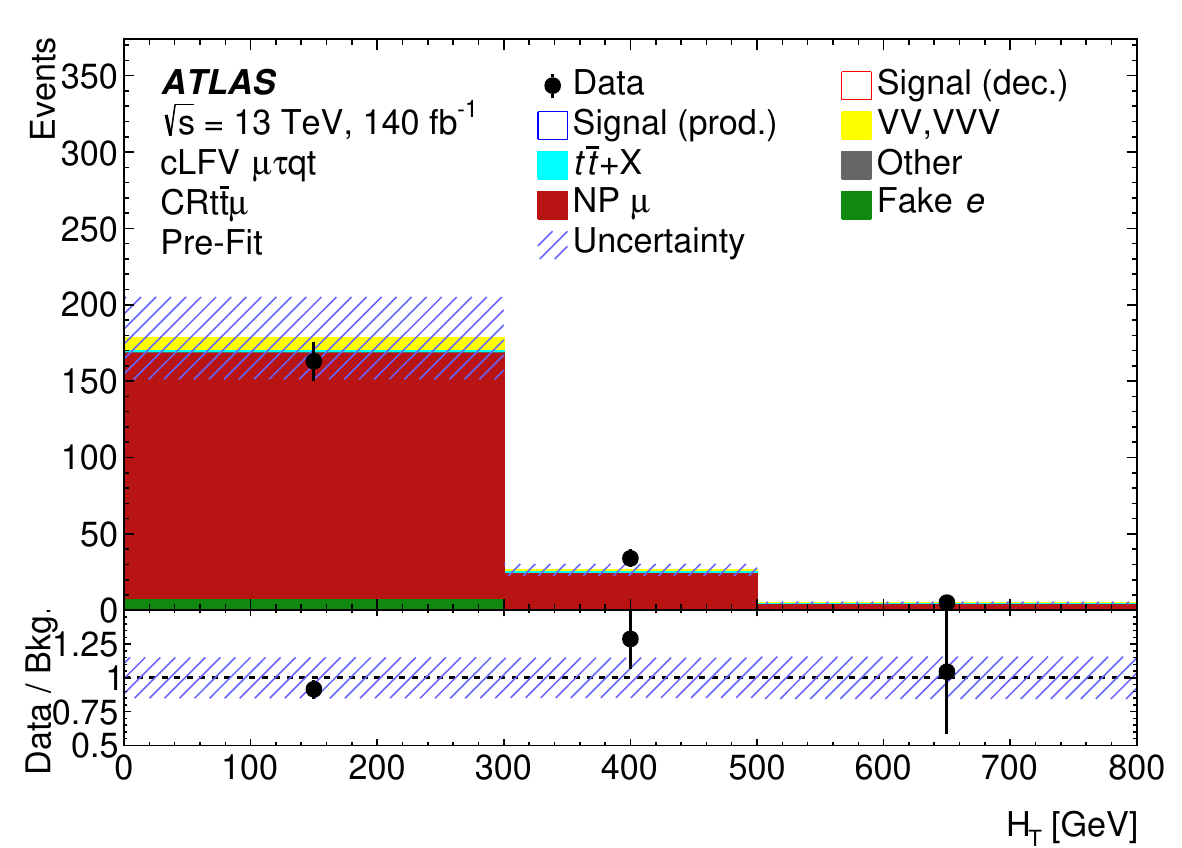}\label{fig:mutau_CRmu_prefit_plot}}
\subfloat[]{\includegraphics[width=0.50\textwidth]{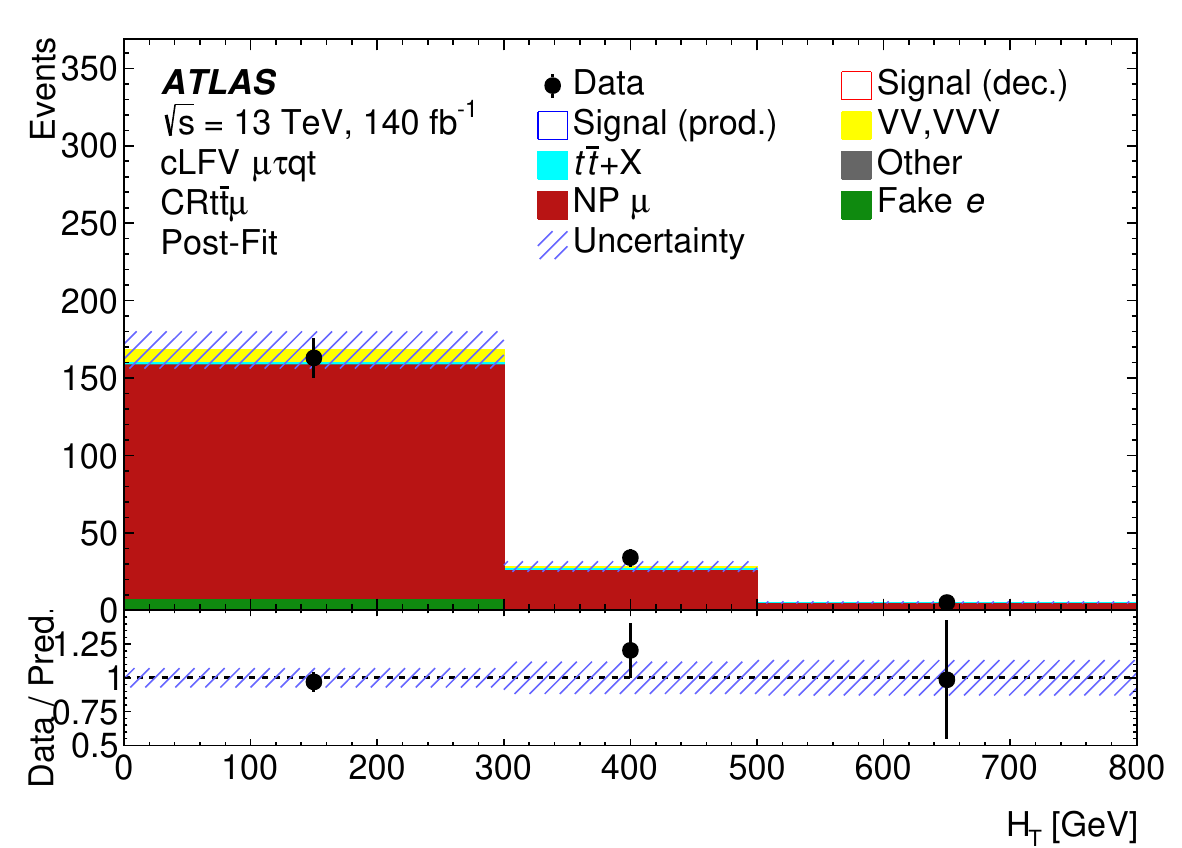}\label{fig:mutau_CRmu_postfit_plot}}\\
\subfloat[]{\includegraphics[width=0.50\textwidth]{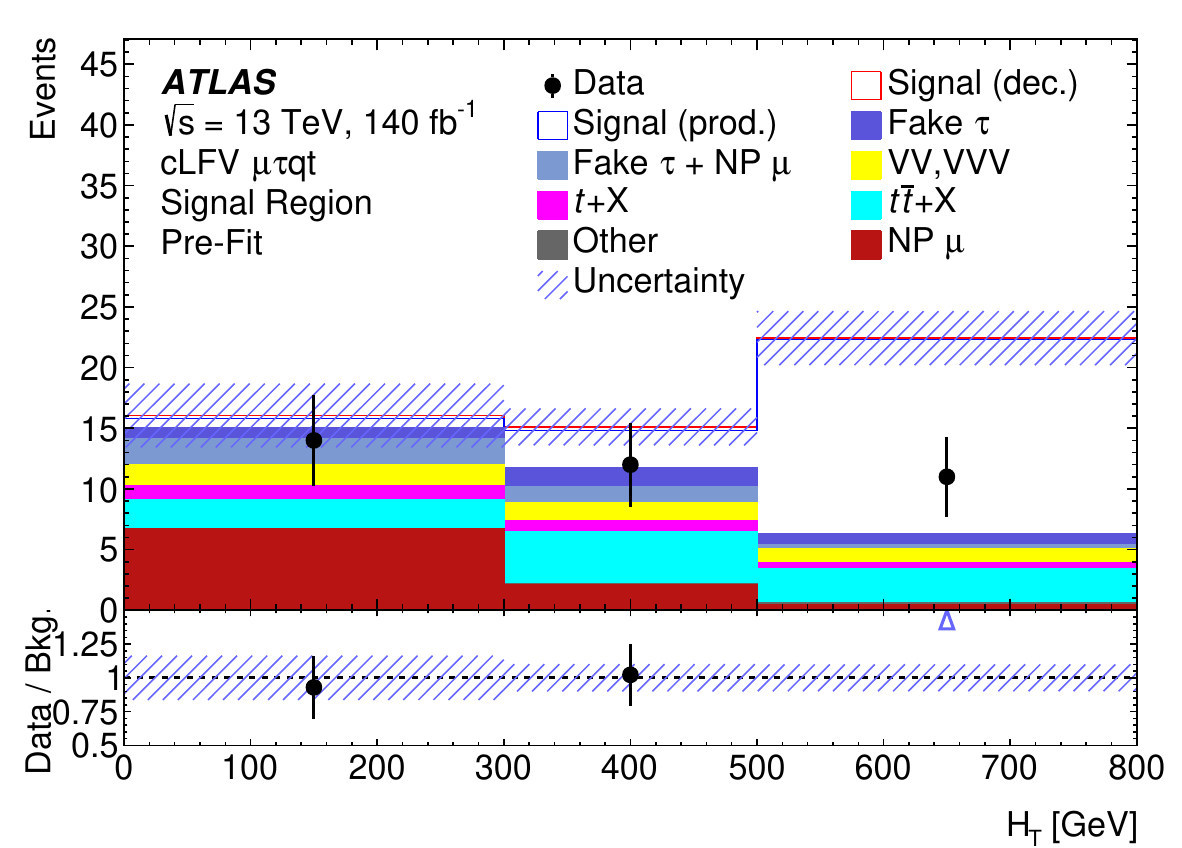}\label{fig:mutau_sr_prefit_plot}}
\subfloat[]{\includegraphics[width=0.50\textwidth]{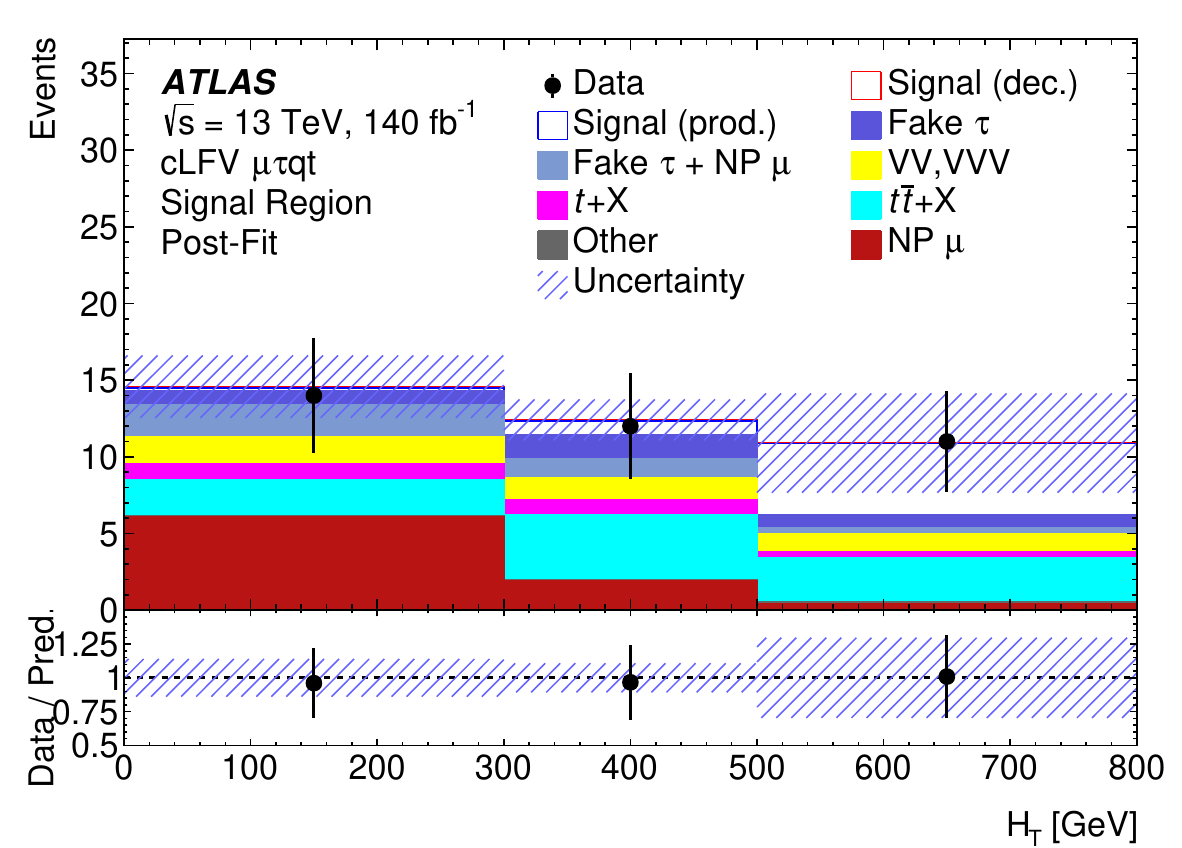}\label{fig:mutau_sr_postfit_plot}}
}
\caption{Observed event yields in $\textrm{CR}\ttbar\mu$ and the SR compared with \protect\subref{fig:mutau_CRmu_prefit_plot}, \protect\subref{fig:mutau_sr_prefit_plot} pre-fit and \protect\subref{fig:mutau_CRmu_postfit_plot}, \protect\subref{fig:mutau_sr_postfit_plot} post-fit expectations from Monte Carlo simulations respectively, as a function of the scalar sum of lepton and jet transverse momenta, $H_{\mathrm{T}}$. The pre-fit signal yield represents all Wilson coefficients set to 0.1 simultaneously for a new physics scale of $\Lambda=1$~\TeV\@. `Signal (prod.)' and `Signal (dec.)' refer to the single-top-quark production and top-quark pair decay signal contributions, respectively, which are together stacked on the background distribution. 
The legend entry of $\ttbar+X$ is the sum of $\ttbar W, \ttbar Z$ and $\ttbar H$.
The lower panels show the ratio of the data (`Data') to the background prediction (`Bkg.') for \protect\subref{fig:mutau_CRmu_prefit_plot}, \protect\subref{fig:mutau_sr_prefit_plot} and to the
background plus signal predictions (`Pred.') for \protect\subref{fig:mutau_CRmu_postfit_plot}, \protect\subref{fig:mutau_sr_postfit_plot}.
The uncertainty band includes both the statistical and systematic uncertainties in the signal and background predictions. For the post-fit plot, correlations among uncertainties were taken into account as determined in the fit under a signal+background hypothesis. An arrow on the lower panel indicates a point outside of the axis range. The last bin includes overflow events.}
\end{figure}

Observed event yields in the SR and $\textrm{CR}\ttbar\mu$, compared with pre-fit expectations from MC simulations, are shown in Figures~\ref{fig:mutau_CRmu_prefit_plot} and~\ref{fig:mutau_sr_prefit_plot}.

Figures~\ref{fig:mutau_CRmu_postfit_plot} and~\ref{fig:mutau_sr_postfit_plot} show the corresponding post-fit distributions, with the post-fit yields given in Table~\ref{tab:mutau_region_postfit_yields}. The background contributions from NP muon production are scaled by a normalisation factor estimated to be $k(\text{NP}\mu) = 1.01 \pm 0.18$, determined by the profile-likelihood fit. The data and SM prediction agree within uncertainties, with the p-value $= 0.053$ for the background-only hypothesis. There is a small upward fluctuation in the data in the highest $H_{\mathrm{T}}$ bin, but no significant cLFV contributions are observed.
Upper limits on the signal cross-section are set at the 95\% confidence level (CL) using the $\mathrm{CL_{s}}$ method~\cite{Read:2002hq}. The corresponding limits on the effective coupling parameters are also calculated~\cite{Cowan:2010js}.
 
The observed and expected 95\% CL limits on the effective coupling strengths, which are set individually, are presented in Table~\ref{tab:eft_mutau_exp_EFTlimits}. This table shows for comparison the existing limits of the relevant Wilson coefficients, which come from a reinterpretation of a previous ATLAS
FCNC search in the $tZq$ channel~\cite{TOPQ-2017-06}. The calculation of these limits is performed in Ref.~\cite{Chala:2018agk}. It can be seen from Table~\ref{tab:clfv_xsecs_all} 
that the signal cross-section is dominated by the $gu \rightarrow t \ell{^\pm} \ell^{' \mp}$ process which leads to stronger exclusion limits on $\mu\tau ut$ interactions than $\mu\tau ct$.
These Wilson coefficient limits are converted into limits on the top-decay branching ratios using Eq.~(\ref{eq:master}) and presented in Table~\ref{tab:eft_mutau_exp_EFTlimits_BRs}. Finally, a fit using two inclusive MC samples with all EFT operators activated simultaneously is used to determine an inclusive branching ratio limit, given in Table~\ref{tab:eft_mutau_exp_BRlimits}. This complements searches for cLFV in $e\mu qt$ interactions by the CMS Collaboration~\cite{CMS:2023iul} that set branching-ratio limits ranging from $\mathcal{B}(t \to e\mu u) < 0.12 \times 10^{-7}$ to $\mathcal{B}(t \to e\mu c) < 4.98 \times 10^{-7}$ for the scalar-structure up-quark coupling and tensor-structure charm-quark couplings, respectively, following the definitions in~\cref{tab:EFTop}.
 
\begin{figure}[t!b!]
\centering
{\centering
\subfloat[]{\includegraphics[width=0.48\textwidth]{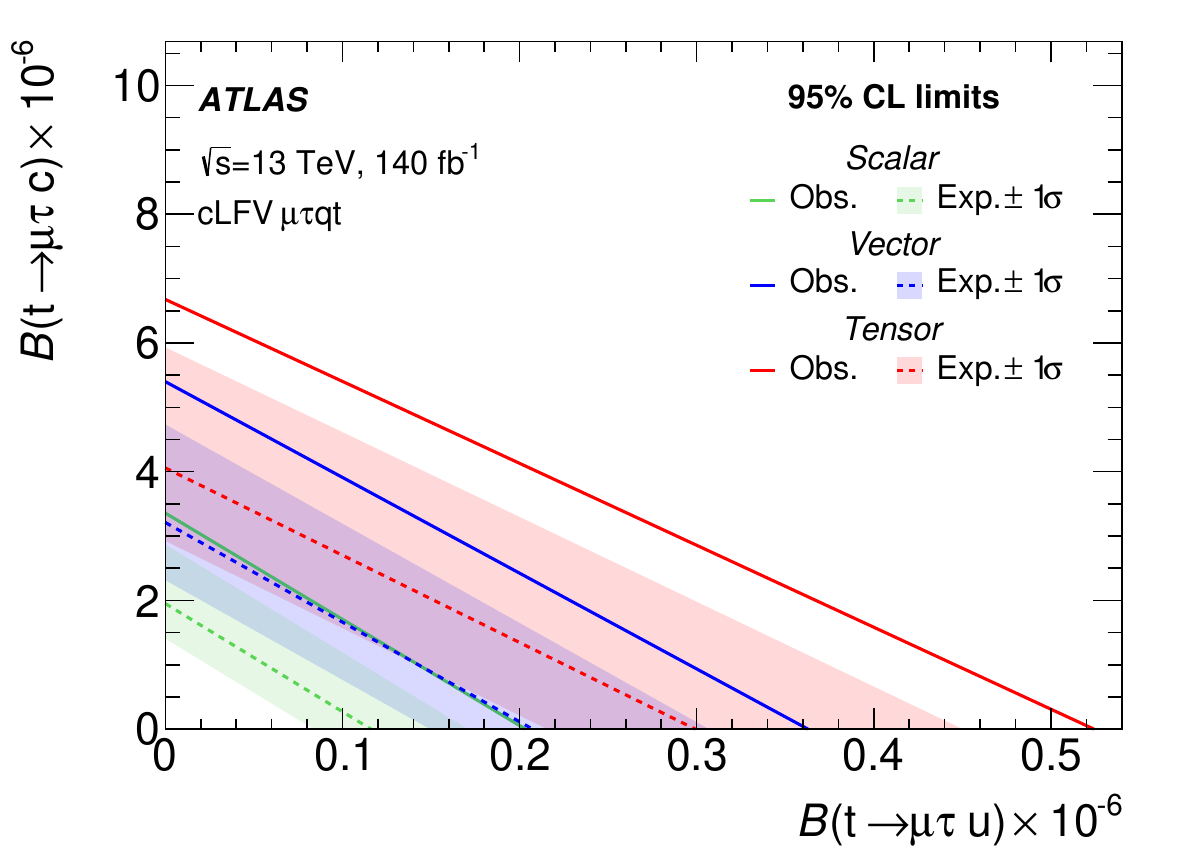}\label{fig:mutau_br_limit_plot_2D}}
\subfloat[]{\includegraphics[width=0.48\textwidth]{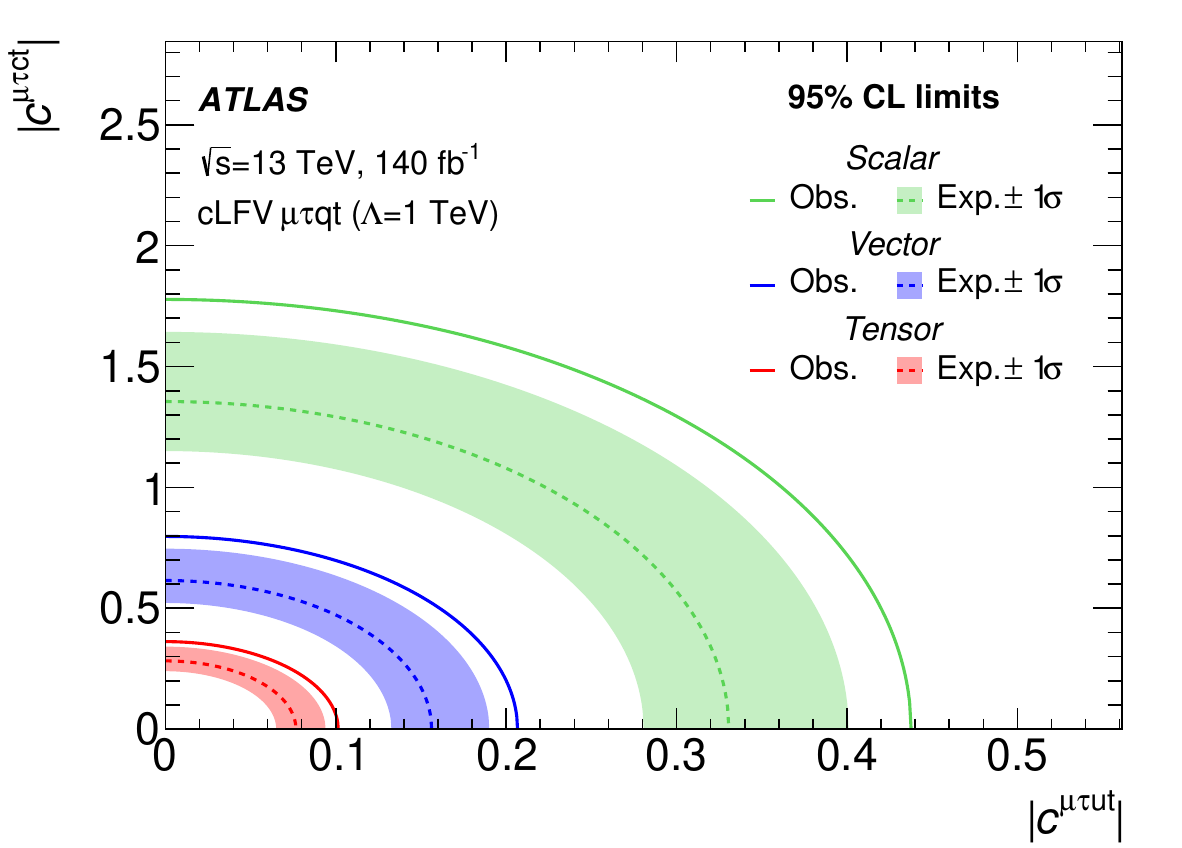}\label{fig:mutau_wc_limit_plot_2D}}
}
\caption{Expected and observed 95\% CL upper limits on the branching ratio corresponding to the decay of a top quark to a muon and a $\tau$-lepton through a cLFV process \protect\subref{fig:mutau_br_limit_plot_2D} and on the relevant Wilson coefficients \protect\subref{fig:mutau_wc_limit_plot_2D} for scalar, vector and tensor couplings, as defined in~\cref{tab:EFTop}. In the vector case, all vector operators are assumed to contribute simultaneously with
the same effective coupling strength. The shaded areas show the $\pm 1$ standard deviation bands around the expected limits.}
\end{figure}

It is also interesting to consider the case in which all vector operators contribute simultaneously with the same effective coupling strength.
This assumes the new physics process couples equally to left- and right-handed fermions.
Figure~\ref{fig:mutau_br_limit_plot_2D} shows the exclusion limits on the branching ratio $\mathcal{B}(t \rightarrow \mu\tau q)$ for $\mu\tau ut$ and $\mu\tau ct$ interactions when considering scalar, vector and tensor couplings.
Figure~\ref{fig:mutau_wc_limit_plot_2D} shows the corresponding limits on the effective coupling strengths under this scenario.
 
The inclusive branching ratio sensitivity is dominated by the cLFV single-top-quark production process, where the \ttbar decay process improves the observed limits by 1\%.
The \ttbar decay process has a more significant impact when considering the charm quark EFT operators, for which it contributes up to 11\% of the sensitivity.
 
In all of the limits extracted, the statistical uncertainty is dominant,
while the largest sources of systematic uncertainty relate to the modelling of the $t\bar{t}+X$ and diboson processes. 
These processes populate the highest $H_{\mathrm{T}}$ bin of the SR, from which most of the sensitivity to the signal process derives. The $\ttbar W$ process utilises a conservative uncertainty in the cross-section, which ranks highly. Additionally, as a non-zero EFT signal contribution is permitted by the fit, the associated signal parton shower uncertainty also plays a leading role in the systematic uncertainty contributions.

\begin{table}[htbp]
\centering
\caption{Post-fit event yields for each analysis region entering the fit, with correlations on the full systematic uncertainties taken into account as determined in the fit under a signal+background hypothesis. The `fake electron' category collects small contributions primarily from \ttgamma~and $Z\gamma$ that enter the event selection due to photon conversions.}
\small
\begin{tabular}{c|S[table-format=2.4]cS[table-format=1.4]|S[table-format=3.3]cS[table-format=2.3]}
\toprule
\textbf{Process} & \multicolumn{3}{c|}{\textbf{SR}} & \multicolumn{3}{c}{\textbf{CR$\ttbar\mu$}} \\
\midrule
\ttbar~+~NP~$\mu$ & 7.9 & $\pm$ & 3.4 & 164 & $\pm$ & 14 \\
\ttW & 3.5 & $\pm$ & 1.8 & 1.2 & $\pm$ & 0.6 \\
\ttH & 3.1 & $\pm$ & 0.4 & 1.26 & $\pm$ & 0.14 \\
\ttZ & 2.9 & $\pm$ & 0.5 & 0.88 & $\pm$ & 0.33 \\
$t$+X & 2.48 &  $\pm$ & 0.18 & & -- & \\
$WZ$ & 3.6 &  $\pm$ & 1.3 & 7.3 & $\pm$ & 2.4 \\
$ZZ$ & 0.59 &  $\pm$ & 0.22 & 1.8 & $\pm$ & 0.6 \\
$VVV$ & 0.01 &  $\pm$ & 0.05 & 0.47 & $\pm$ & 0.24 \\
Fake electron & & -- & & 7 & $\pm$ & 4 \\
Fake~$\tau$ & 3.3 & $\pm$ & 0.4 & & -- & \\
Fake~$\tau$~+~NP~$\mu$ & 3.7 & $\pm$ & 2.7 & & -- & \\
$t$+X~+~NP$~\mu$ & 0.29 & $\pm$ & 0.31 & 15 & $\pm$ & 5 \\
$Z$~+~NP~$\mu$ & 0.192 & $\pm$ & 0.010 & 1.8 & $\pm$ & 1.0 \\
Other NP~$\mu$ & 0.051 & $\pm$ & 0.010 & & -- &  \\
Other & 0.23 & $\pm$ & 0.11 & 1.1 & $\pm$ & 0.6 \\
\midrule
Signal (\ttbar) & 0.19 & $\pm$ & 0.14 & 0.025 & $\pm$ & 0.019 \\
Signal (single-top) & 6 & $\pm$ & 4 & 0.022 & $\pm$ & 0.023 \\
\midrule
Total & 38 & $\pm$ & 5 & 201 & $\pm$ & 14 \\
\midrule
Data & 37 & & & 202 & & \\
\bottomrule
\end{tabular}
\label{tab:mutau_region_postfit_yields}
\end{table}

\begin{table}[htbp]
\centering
\caption{Expected and observed 95\% CL upper limits on Wilson coefficients corresponding to 2Q2L EFT operators that could introduce a cLFV top decay in the \mutau channel, and existing limits from Ref.~\cite{Chala:2018agk} (previous). The previous limits shown for $c^{1(ijk3)}_{\mathsf{lequ}}$ and $c^{3(ijk3)}_{\mathsf{lequ}}$ are tightened by a factor of $\sqrt{2}$ relative to the individual values given in Ref.~\cite{Chala:2018agk}, as that interpretation does not apply the assumption that these operators are Hermitian. The assumption of Hermitian operators results in a larger cross section for a Wilson coefficient of a given value, as its Hermitian conjugate is also considered. Consequently, an observed signal strength results in a tighter upper limit on a Wilson coefficient than if its Hermitian conjugate were considered separately. Results are shown separately for the $\mu\tau ut$ and $\mu\tau ct$ interactions. The lepton generations are denoted by $i,j=2,3$ for $\mu$ and $\tau$ (where $i \neq j$) and the quark generations are denoted by $k=1,2$ for $u$ and $c$, respectively.}
\small
\renewcommand{\arraystretch}{1.2}
\begin{tabular}{c|cccccc}
\toprule
& \multicolumn{6}{c}{\textbf{95\% CL upper limits on $|c| / \Lambda^{2}$ [$\textrm{\TeV}^{-2}$] }} \\[0.5ex]
& $c^{-(ijk3)}_{\mathsf{lq}}$ & $c^{(ijk3)}_{\mathsf{eq}}$ & $c^{(ijk3)}_{\mathsf{lu}}$ & $c^{(ijk3)}_{\mathsf{eu}}$ & $c^{1(ijk3)}_{\mathsf{lequ}}$ & $c^{3(ijk3)}_{\mathsf{lequ}}$ \\[1.0ex] \hline
\textbf{Previous (u)}~\cite{Chala:2018agk} & 12 & 12 & 12 & 12 & 18 & 2.4 \\
\textbf{Expected (u)} & 0.33 & 0.31 & 0.3 & 0.32 & 0.33 & 0.08 \\
\textbf{Observed (u)} & 0.43 & 0.41 & 0.4 & 0.42 & 0.44 & 0.10 \\
\midrule
\textbf{Previous (c)}~\cite{Chala:2018agk} & 14 & 14 & 14 & 14 & 21 & 2.6 \\
\textbf{Expected (c)} & 1.3 & 1.2 & 1.2 & 1.2 & 1.4 & 0.28 \\
\textbf{Observed (c)} & 1.6 & 1.6 & 1.6 & 1.6 & 1.8 & 0.36 \\
\bottomrule
\end{tabular}
\label{tab:eft_mutau_exp_EFTlimits}
\end{table}

\begin{table}[htbp]
\centering
\caption{Expected and observed 95\% CL upper limits on the branching ratio ($\mathcal{B}$) corresponding to the decay of a top quark to a muon and a $\tau$-lepton through a cLFV process using specific Wilson coefficients corresponding to 2Q2L EFT operators. Results are shown separately for $\mu\tau ut$ and $\mu\tau ct$ interactions. The lepton generations are denoted by $i,j=2,3$ for $\mu$ and $\tau$ (where $i \neq j$) and the quark generations are denoted by $k=1,2$ for $u$ and $c$, respectively.}
\small
\renewcommand{\arraystretch}{1.2}
\begin{tabular}{c|cccccc}
\toprule
& \multicolumn{6}{c}{\textbf{95\% CL upper limits on $\mathcal{B}(t \rightarrow \mu\tau q) \quad (\times ~ 10^{-7})$}} \\[0.5ex]
& $c^{-(ijk3)}_{\mathsf{lq}}$ & $c^{(ijk3)}_{\mathsf{eq}}$ & $c^{(ijk3)}_{\mathsf{lu}}$ & $c^{(ijk3)}_{\mathsf{eu}}$ & $c^{1(ijk3)}_{\mathsf{lequ}}$ & $c^{3(ijk3)}_{\mathsf{lequ}}$ \\[1.0ex] \hline
\textbf{Expected (u)} & 2.3 & 2.0 & 1.9 & 2.2 & 1.2 & 3.0 \\
\textbf{Observed (u)} & 4.0 & 3.6 & 3.3 & 3.8 & 2.0 & 5.2 \\
\midrule
\textbf{Expected (c)} & 33 & 32 & 32 & 33 & 20 & 41 \\
\textbf{Observed (c)} & 56 & 54 & 53 & 54 & 34 & 67 \\
\bottomrule
\end{tabular}
\label{tab:eft_mutau_exp_EFTlimits_BRs}
\end{table}
 
\begin{table}[htbp]
\setlength\extrarowheight{5pt}
\centering
\caption{Expected and observed 95\% CL upper limits on the inclusive branching ratio ($\mathcal{B}$) corresponding to the decay of a top quark to a muon and a $\tau$-lepton 
through a cLFV process. Limits are shown for the statistical uncertainty only and for the full set of statistical and systematic uncertainties.}
\small
\begin{tabular}{c|c|c}
\toprule
& \multicolumn{2}{c}{\textbf{95\% CL upper limits on $\mathcal{B}(t \rightarrow \mu\tau q)$}} \\
& ~~~~~~\textbf{Stat. uncertainty}~~~~~~ & \textbf{Stat.+syst. uncertainties} \\
\hline
\textbf{Expected} & $4.6 \times 10^{-7}$ & $5.0 \times 10^{-7}$ \\
\textbf{Observed} & $8.2 \times 10^{-7}$ & $8.7 \times 10^{-7}$ \\
\bottomrule
\end{tabular}
\label{tab:eft_mutau_exp_BRlimits}
\end{table}
 
% The next lines are included from the .//sections/LeptoquarkInterpretation.tex input file
\subsection{Leptoquark interpretation}
\label{sec:LQSearch}
 
A search for the production of a scalar leptoquark $S_{1}$ was performed using
the same analysis strategy as
the EFT interpretation described in Section~\ref{sec:eft_search}, based on the $H_{\mathrm{T}}$ distribution. 
As discussed in Section~\ref{sec:DataMC}, MC samples are generated in which the $S_{1}$ leptoquark is produced in association with a charged lepton via quark-gluon fusion and subsequently decays into a top quark and a second charged lepton. The samples are simulated with masses in the range of $0.5 < m_{S_{1}} < 2.5$ \TeV in steps of $250$ \GeV, and coupling values $\lambda^{\textrm{LQ}}$ between $0.5$ and $3.5$ in steps of $0.5$, using the generational hierarchy shown in Eq.~(\ref{eq:lqhierarchy}). Observed and expected limits are set on the $S_{1}$ cross-section using a profile-likelihood ratio in the asymptotic limit with nuisance parameters to account for systematic uncertainties. The limits on the cross-section were extracted at the $95\%$ CL\@.
All experimental and modelling uncertainties that were considered for the EFT interpretation were also considered as nuisance parameters in the leptoquark fit, but with the free-floating parameters representing the Wilson coefficients replaced by a free-floating parameter for the cross-section of $S_{1}$ production.
The same background estimation techniques, event selection and region definitions were used. The leptoquark fit was found to have comparable sensitivity to the background modelling as for the EFT interpretation, where the largest sources of systematic uncertainty were the $t\bar{t}W$ modelling and $WZ$ normalisation. The sensitivity is dominated by the statistical uncertainty.
 
\begin{figure}[b!tp]
\centering
{\centering
\subfloat[]{\includegraphics[width=0.50\textwidth]{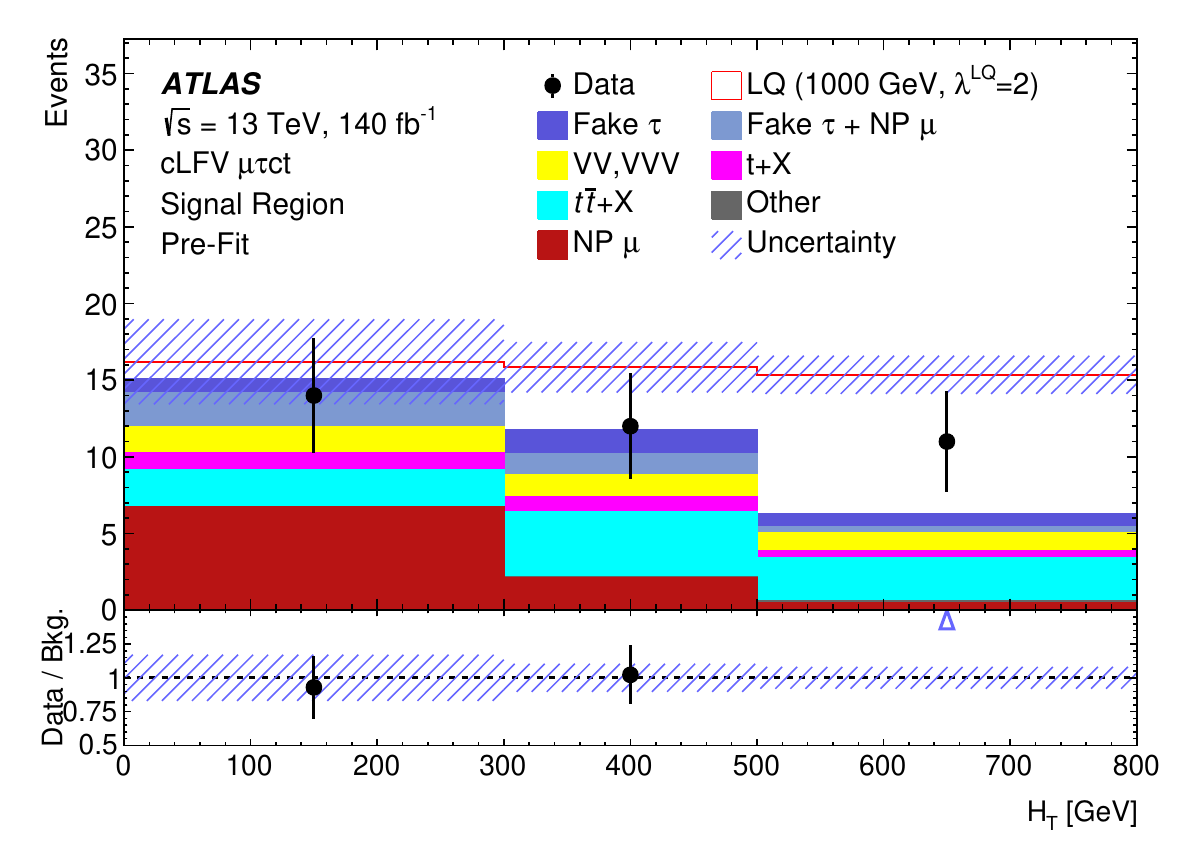}\label{fig:lq_prefit_plot_kappa2}}
\subfloat[]{\includegraphics[width=0.50\textwidth]{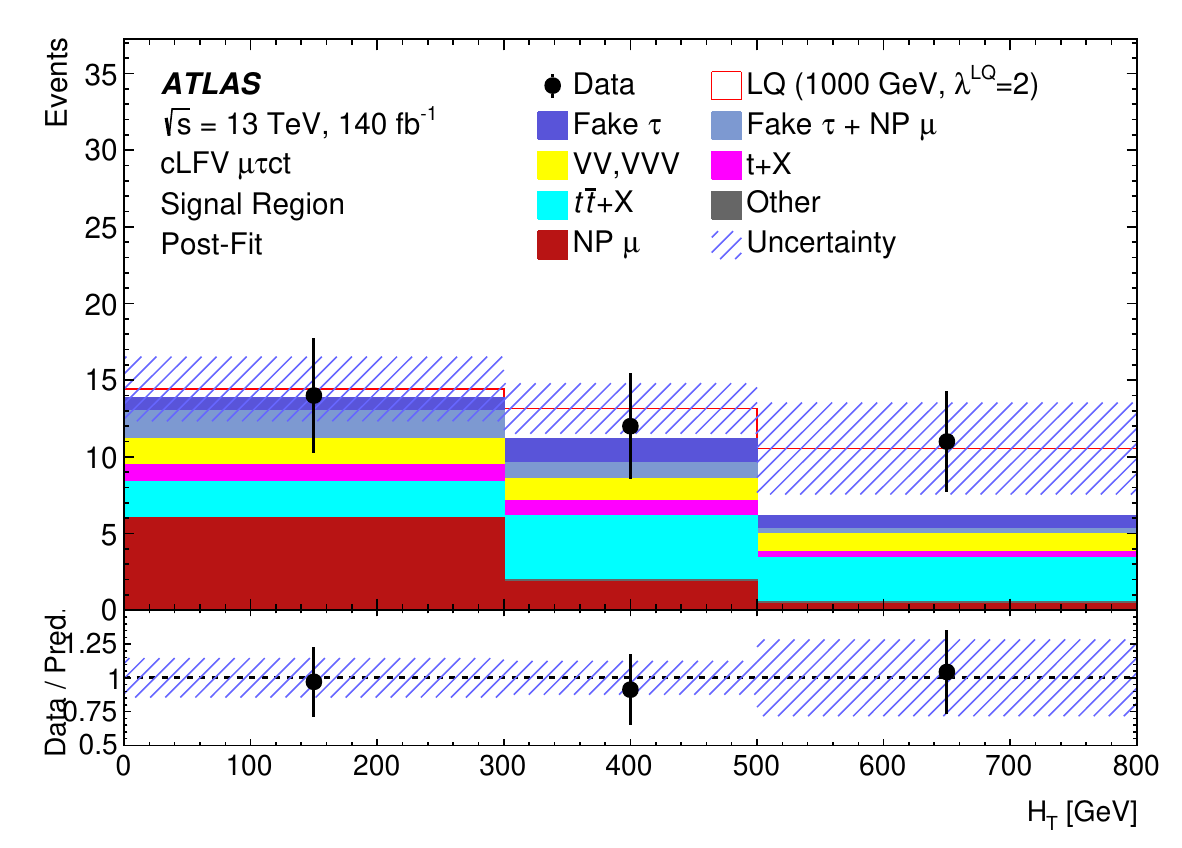}\label{fig:lq_prefit_plot_kappa2_post}}
}
\caption{Observed event yields in the SR compared with \protect\subref{fig:lq_prefit_plot_kappa2} pre-fit and \protect\subref{fig:lq_prefit_plot_kappa2_post} post-fit expectations from Monte Carlo simulations. The signal yields represent a leptoquark mass of $m_{S_{1}} = 1$ \TeV and a coupling strength of $\lambda^{\textrm{LQ}} = 2.0$. The uncertainty band includes both the statistical and systematic uncertainties in the signal and background predictions. 
The legend entry of $\ttbar+X$ is the sum of $\ttbar W, \ttbar Z$ and $\ttbar H$.
The lower panels show the ratio of the data (`Data') to the background prediction (`Bkg.') for \protect\subref{fig:lq_prefit_plot_kappa2}, and to the background plus signal predictions (`Pred.') for \protect\subref{fig:lq_prefit_plot_kappa2_post}.
An arrow on the lower panel indicates a point outside of the axis range. The last bin includes overflow events.}
\label{fig:lq_prefit_plot}
\end{figure}

\begin{figure}[t!bp]
\centering
{\centering
\subfloat[]{\includegraphics[width=0.50\textwidth]{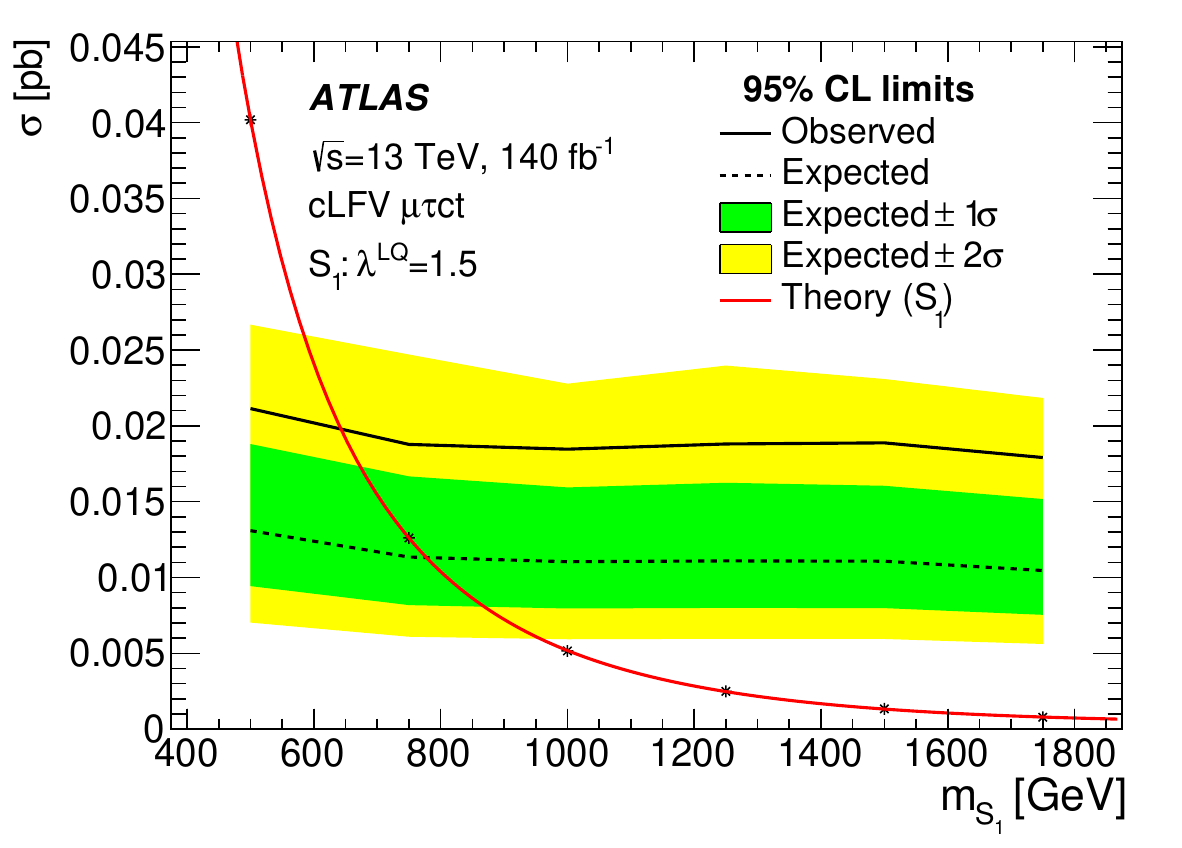}\label{fig:lq_1d_limit_lam1p5}}
\subfloat[]{\includegraphics[width=0.50\textwidth]{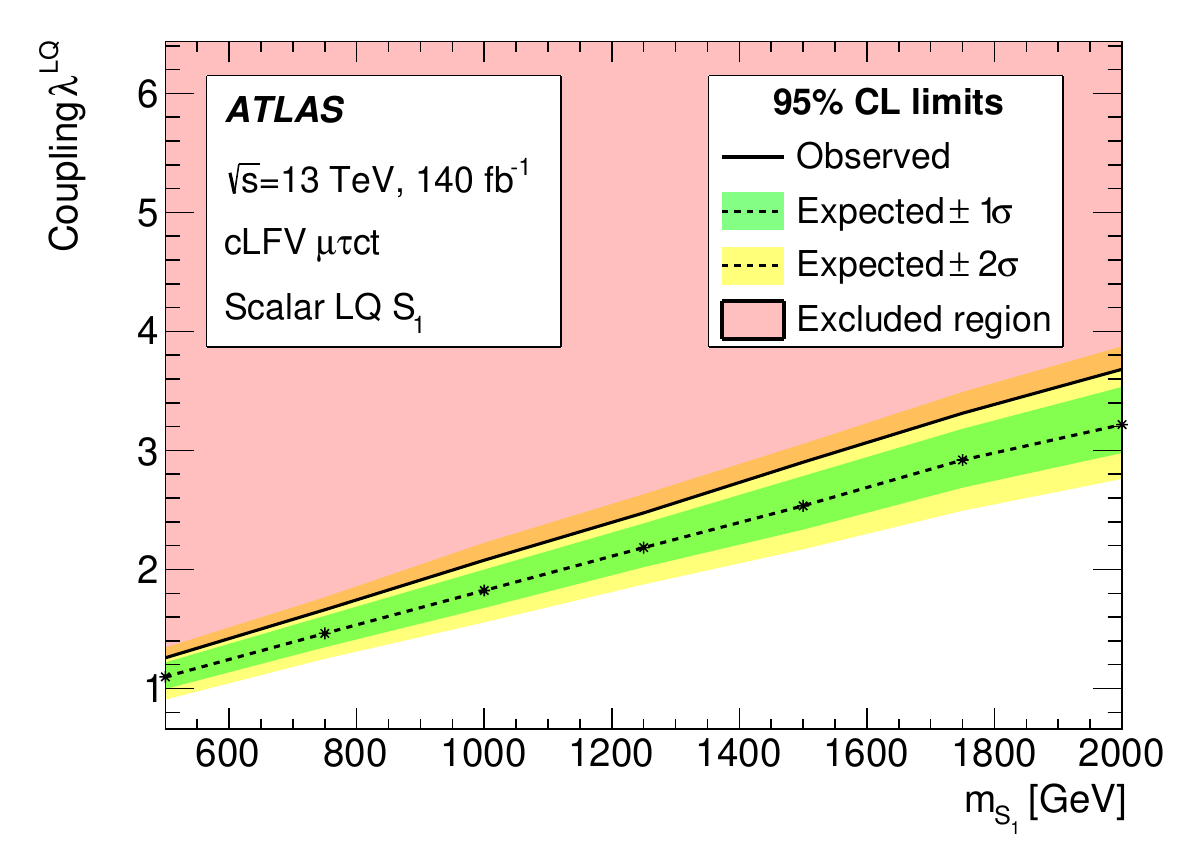}\label{fig:lq_2d_limit}}
}
\caption{\protect\subref{fig:lq_1d_limit_lam1p5} Expected and observed 95\% CL upper limits on the production cross-section of a scalar leptoquark with $\lambda^{\textrm{LQ}} = 1.5$ and for a range of mass values. The overlaid theory line shows a polynominal fit to the theoretical cross-section values. \protect\subref{fig:lq_2d_limit} Expected and observed 95\% CL upper limits as a function of the leptoquark mass $m_{S_{1}}$ and the coupling strength $\lambda^{\textrm{LQ}}$. The solid line shows the observed limit while the dashed line shows the expected limit with the $\pm 1$ and $\pm 2$ standard deviation bands. The shaded area above the observed limit represents the excluded region. }
\end{figure}

Figure~\ref{fig:lq_prefit_plot} shows the comparison between the data and both the pre-fit and post-fit expectations for the $H_{\mathrm{T}}$ distribution in the SR, for the $S_{1}$ model at $m_{S_{1}} = 1$ \TeV and  $\lambda^{\textrm{LQ}} = 2.0$. Example cross-section limits for different $m_{S_{1}}$ samples are shown in Figure~\ref{fig:lq_1d_limit_lam1p5} for $\lambda^{\textrm{LQ}} = 1.5$. The experimentally observed (expected) cross-section limit is approximately $17-21$~fb ($10-13$~fb) depending on the leptoquark mass.
 
Figure~\ref{fig:lq_2d_limit} shows the observed and expected limits as a function of both $m_{S_{1}}$ and $\lambda^{\textrm{LQ}}$. Observed (expected) upper limits on leptoquark coupling strengths are set at the $95\%$ CL, ranging from $\lambda^{\textrm{LQ}} = 1.3~(1.1)$ to $\lambda^{\textrm{LQ}} = 3.7~(3.2)$ for masses between $0.5$ and $2.0$ \TeV.
The sensitivity for $m_{S_{1}} > 2.0$ \TeV was not high enough to set $95\%$ CL limits for the generated values of $\lambda^{\textrm{LQ}}$.

% End of text imported from the .//sections/LeptoquarkInterpretation.tex input file

% End of text imported from the .//sections/Results.tex input file

\FloatBarrier

% The next lines are included from the .//sections/Conclusion.tex input file
\section{Conclusions}
A search for charged-lepton-flavour violation in top-quark interactions is presented, using proton–proton collision data collected by the ATLAS experiment at the LHC at a centre-of-mass energy of $\sqrt{s} = 13$ \TeV, corresponding to an integrated luminosity of $140$~\ifb{}.  The cLFV single-top-quark production and \ttbar decay processes, $gq_{k} \to t\mu\tau$ and $\ttbar \to (\ell\ell^{'}q)((W\to \ell\nu) b)$ respectively, were considered, whereby the SM top quark always produces a muon in the final state.
The signal topology is thus characterised by the presence of two isolated muons, one hadronically decaying $\tau$-lepton, at least one jet and exactly one $b$-tagged jet; to suppress backgrounds the muons are required to have the same electric charge.

The SM prediction is found to agree with the data to within $1.6\sigma$.
The limits obtained on the Wilson coefficients improve upon the previous results from a reinterpretation of an ATLAS FCNC $tZq$ analysis, from a factor of 7.2 for $|c^{3(2323)}_{\mathsf{lequ}}| / \Lambda^{2}$  (for $\mu\tau ct$) to a factor of 41 for $|c^{1(2313)}_{\mathsf{lequ}}| / \Lambda^{2}$ (for $\mu\tau ut$). These Wilson coefficient limits are translated into an observed 95\% confidence level upper limit on the $t \to \mu\tau q$ decay branching ratio of $\mathcal{B}(t \to \mu\tau q) < 8.7 \times 10^{-7}$.

A search for single scalar leptoquark $S_{1}$ production was also performed assuming fixed hierarchical couplings to multiple quark and lepton flavour generations. Limits were set on this model as a function of the leptoquark mass $m_{S_{1}}$ and an overall coupling strength $\lambda^{\textrm{LQ}}$. Upper limits on leptoquark coupling strengths are set at the $95\%$ CL, ranging from $\lambda^{\textrm{LQ}} = 1.3$ to $\lambda^{\textrm{LQ}} = 3.7$ for masses between $0.5$ and $2.0$ \TeV.
 
\clearpage

% End of text imported from the .//sections/Conclusion.tex input file

\section*{Acknowledgements}

% The next lines are included from the .//acknowledgements/Acknowledgements.tex input file

We thank CERN for the very successful operation of the LHC and its injectors, as well as the support staff at
CERN and at our institutions worldwide without whom ATLAS could not be operated efficiently.
 
The crucial computing support from all WLCG partners is acknowledged gratefully, in particular from CERN, the ATLAS Tier-1 facilities at TRIUMF/SFU (Canada), NDGF (Denmark, Norway, Sweden), CC-IN2P3 (France), KIT/GridKA (Germany), INFN-CNAF (Italy), NL-T1 (Netherlands), PIC (Spain), RAL (UK) and BNL (USA), the Tier-2 facilities worldwide and large non-WLCG resource providers. Major contributors of computing resources are listed in Ref.~\cite{ATL-SOFT-PUB-2023-001}.
 
We gratefully acknowledge the support of ANPCyT, Argentina; YerPhI, Armenia; ARC, Australia; BMWFW and FWF, Austria; ANAS, Azerbaijan; CNPq and FAPESP, Brazil; NSERC, NRC and CFI, Canada; CERN; ANID, Chile; CAS, MOST and NSFC, China; Minciencias, Colombia; MEYS CR, Czech Republic; DNRF and DNSRC, Denmark; IN2P3-CNRS and CEA-DRF/IRFU, France; SRNSFG, Georgia; BMBF, HGF and MPG, Germany; GSRI, Greece; RGC and Hong Kong SAR, China; ISF and Benoziyo Center, Israel; INFN, Italy; MEXT and JSPS, Japan; CNRST, Morocco; NWO, Netherlands; RCN, Norway; MNiSW, Poland; FCT, Portugal; MNE/IFA, Romania; MESTD, Serbia; MSSR, Slovakia; ARRS and MIZ\v{S}, Slovenia; DSI/NRF, South Africa; MICINN, Spain; SRC and Wallenberg Foundation, Sweden; SERI, SNSF and Cantons of Bern and Geneva, Switzerland; MOST, Taipei; TENMAK, T\"urkiye; STFC, United Kingdom; DOE and NSF, United States of America.
 
Individual groups and members have received support from BCKDF, CANARIE, CRC and DRAC, Canada; CERN-CZ, PRIMUS 21/SCI/017 and UNCE SCI/013, Czech Republic; COST, ERC, ERDF, Horizon 2020, ICSC-NextGenerationEU and Marie Sk{\l}odowska-Curie Actions, European Union; Investissements d'Avenir Labex, Investissements d'Avenir Idex and ANR, France; DFG and AvH Foundation, Germany; Herakleitos, Thales and Aristeia programmes co-financed by EU-ESF and the Greek NSRF, Greece; BSF-NSF and MINERVA, Israel; Norwegian Financial Mechanism 2014-2021, Norway; NCN and NAWA, Poland; La Caixa Banking Foundation, CERCA Programme Generalitat de Catalunya and PROMETEO and GenT Programmes Generalitat Valenciana, Spain; G\"{o}ran Gustafssons Stiftelse, Sweden; The Royal Society and Leverhulme Trust, United Kingdom.
 
In addition, individual members wish to acknowledge support from CERN: European Organization for Nuclear Research (CERN PJAS); Chile: Agencia Nacional de Investigaci\'on y Desarrollo (FONDECYT 1190886, FONDECYT 1210400, FONDECYT 1230812, FONDECYT 1230987); China: National Natural Science Foundation of China (NSFC - 12175119, NSFC 12275265, NSFC-12075060); Czech Republic: PRIMUS Research Programme (PRIMUS/21/SCI/017); European Union: European Research Council (ERC - 948254, ERC 101089007), Horizon 2020 Framework Programme (MUCCA - CHIST-ERA-19-XAI-00), European Union, Future Artificial Intelligence Research (FAIR-NextGenerationEU PE00000013), Italian Center for High Performance Computing, Big Data and Quantum Computing (ICSC, NextGenerationEU); France: Agence Nationale de la Recherche (ANR-20-CE31-0013, ANR-21-CE31-0013, ANR-21-CE31-0022), Investissements d'Avenir Labex (ANR-11-LABX-0012); Germany: Baden-Württemberg Stiftung (BW Stiftung-Postdoc Eliteprogramme), Deutsche Forschungsgemeinschaft (DFG - 469666862, DFG - CR 312/5-2); Italy: Istituto Nazionale di Fisica Nucleare (ICSC, NextGenerationEU); Japan: Japan Society for the Promotion of Science (JSPS KAKENHI Grant No. 22KK0227, JSPS KAKENHI JP21H05085, JSPS KAKENHI JP22H01227, JSPS KAKENHI JP22H04944); Netherlands: Netherlands Organisation for Scientific Research (NWO Veni 2020 - VI.Veni.202.179); Norway: Research Council of Norway (RCN-314472); Poland: Polish National Agency for Academic Exchange (PPN/PPO/2020/1/00002/U/00001), Polish National Science Centre (NCN 2021/42/E/ST2/00350, NCN OPUS nr 2022/47/B/ST2/03059, NCN UMO-2019/34/E/ST2/00393, UMO-2020/37/B/ST2/01043, UMO-2021/40/C/ST2/00187, UMO-2022/47/O/ST2/00148); Slovenia: Slovenian Research Agency (ARIS grant J1-3010); Spain: BBVA Foundation (LEO22-1-603), Generalitat Valenciana (Artemisa, FEDER, IDIFEDER/2018/048), Ministry of Science and Innovation (MCIN \& NextGenEU -PCI2022-135018-2, MICIN \& FEDER -PID2021-125273NB, RYC2019-028510-I, RYC2020-030254-I, RYC2021-031273-I, RYC2022-038164-I), PROMETEO and GenT Programmes Generalitat Valenciana (CIDEGENT/2019/023, CIDEGENT/2019/027); Sweden: Swedish Research Council (VR 2018-00482, VR 2022-03845, VR 2022-04683, VR grant 2021-03651), Knut and Alice Wallenberg Foundation (KAW 2017.0100, KAW 2018.0157, KAW 2018.0458, KAW 2019.0447, KAW 2022.0358); Switzerland: Swiss National Science Foundation (SNSF - PCEFP2\_194658); United Kingdom: Leverhulme Trust (Leverhulme Trust RPG-2020-004), Royal Society (NIF-R1-231091); United States of America: Neubauer Family Foundation.

% End of text imported from the .//acknowledgements/Acknowledgements.tex input file

\clearpage

\printbibliography
 
\clearpage
% ATLAS Collaboration author list
% Reference date of TOPQ-2023-23 is 2023-12-14
% Author list last updated on date 15-MAY-24
% Data extracted on 15-May-2024 for paper reference TOPQ-2023-23
% at 1:47pm
 
\begin{flushleft}
\hypersetup{urlcolor=black}
{\Large The ATLAS Collaboration}

\bigskip

\AtlasOrcid[0000-0002-6665-4934]{G.~Aad}$^\textrm{\scriptsize 103}$,
\AtlasOrcid[0000-0001-7616-1554]{E.~Aakvaag}$^\textrm{\scriptsize 16}$,
\AtlasOrcid[0000-0002-5888-2734]{B.~Abbott}$^\textrm{\scriptsize 121}$,
\AtlasOrcid[0000-0002-1002-1652]{K.~Abeling}$^\textrm{\scriptsize 55}$,
\AtlasOrcid[0000-0001-5763-2760]{N.J.~Abicht}$^\textrm{\scriptsize 49}$,
\AtlasOrcid[0000-0002-8496-9294]{S.H.~Abidi}$^\textrm{\scriptsize 29}$,
\AtlasOrcid[0009-0003-6578-220X]{M.~Aboelela}$^\textrm{\scriptsize 44}$,
\AtlasOrcid[0000-0002-9987-2292]{A.~Aboulhorma}$^\textrm{\scriptsize 35e}$,
\AtlasOrcid[0000-0001-5329-6640]{H.~Abramowicz}$^\textrm{\scriptsize 152}$,
\AtlasOrcid[0000-0002-1599-2896]{H.~Abreu}$^\textrm{\scriptsize 151}$
% ...
\bigskip
\\

% ...
$^{16}$Department for Physics and Technology, University of Bergen, Bergen; Norway.\\
% ...
$^{29}$Physics Department, Brookhaven National Laboratory, Upton NY; United States of America.\\
% ...
$^{35}$$^{(a)}$Facult\'e des Sciences Ain Chock, R\'eseau Universitaire de Physique des Hautes Energies - Universit\'e Hassan II, Casablanca;$^{(b)}$Facult\'{e} des Sciences, Universit\'{e} Ibn-Tofail, K\'{e}nitra;$^{(c)}$Facult\'e des Sciences Semlalia, Universit\'e Cadi Ayyad, LPHEA-Marrakech;$^{(d)}$LPMR, Facult\'e des Sciences, Universit\'e Mohamed Premier, Oujda;$^{(e)}$Facult\'e des sciences, Universit\'e Mohammed V, Rabat;$^{(f)}$Institute of Applied Physics, Mohammed VI Polytechnic University, Ben Guerir; Morocco.\\
% ...
$^{44}$Physics Department, Southern Methodist University, Dallas TX; United States of America.\\
% ...
$^{49}$Fakult\"{a}t Physik , Technische Universit{\"a}t Dortmund, Dortmund; Germany.\\
% ...
$^{55}$II. Physikalisches Institut, Georg-August-Universit\"{a}t G\"ottingen, G\"ottingen; Germany.\\
% ...
$^{103}$CPPM, Aix-Marseille Universit\'e, CNRS/IN2P3, Marseille; France.\\
% ...
$^{121}$Homer L. Dodge Department of Physics and Astronomy, University of Oklahoma, Norman OK; United States of America.\\
% ...
$^{151}$Department of Physics, Technion, Israel Institute of Technology, Haifa; Israel.\\
$^{152}$Raymond and Beverly Sackler School of Physics and Astronomy, Tel Aviv University, Tel Aviv; Israel.\\
% ...
$^{a}$ Also Affiliated with an institute covered by a cooperation agreement with CERN.\\
$^{b}$ Also at An-Najah National University, Nablus; Palestine.\\
$^{c}$ Also at Borough of Manhattan Community College, City University of New York, New York NY; United States of America.\\
$^{d}$ Also at Center for High Energy Physics, Peking University; China.\\
$^{e}$ Also at Center for Interdisciplinary Research and Innovation (CIRI-AUTH), Thessaloniki; Greece.\\
$^{f}$ Also at Centro Studi e Ricerche Enrico Fermi; Italy.\\
% ...

\end{flushleft}

% Created with Glance <Atlas.Glance@cern.ch>

\end{document}